\documentclass[journal]{IEEEtran}

\usepackage[font=footnotesize,labelfont=sf,textfont=sf]{caption}
\usepackage{url}
\usepackage{graphicx}
\usepackage{multirow}
\usepackage{subcaption}
\usepackage{xspace}
\graphicspath{{figures/}}
\usepackage{mathtools,commath,amsmath,amssymb,amsfonts,bm}\interdisplaylinepenalty=2500
\usepackage{tabularx,adjustbox}
\newcolumntype{L}[1]{>{\hsize=#1\hsize\raggedright\arraybackslash}X}%
\newcolumntype{R}[1]{>{\hsize=#1\hsize\raggedleft\arraybackslash}X}%
\newcolumntype{C}[1]{>{\hsize=#1\hsize\centering\arraybackslash}X}%
\usepackage{todonotes}
\usepackage{breqn}
\clearpage
\extrafloats{100}
\RequirePackage{tcolorbox}
\tcbuselibrary{breakable}
\tcbset{parbox=false, left=1ex, right=1ex}

\usepackage{makecell}

\def\vec#1{\mathchoice{\mbox{\boldmath$\displaystyle#1$}}
  {\mbox{\boldmath$\textstyle#1$}}
  {\mbox{\boldmath$\scriptstyle#1$}}
  {\mbox{\boldmath$\scriptscriptstyle#1$}}}

\newcommand{\ie}{\textit{i.e.},\xspace}
\newcommand{\etal}{\textit{et~al.}\xspace}
\newcommand{\eg}{\textit{e.g.},\xspace}
\newcommand{\mech}{\ensuremath{\mathcal{M}}\xspace}
 \newtheorem{definition}{Definition}[section]

\begin{document}
\title{Privacy and Robustness in Federated Learning: Attacks and Defenses}

\author{Lingjuan~Lyu$^{*}$,
        Han~Yu$^{*}$,
        Xingjun~Ma, Chen~Chen, Lichao~Sun, Jun~Zhao,
        Qiang~Yang$^{*}$,~\IEEEmembership{Fellow,~IEEE}, and Philip~S.~Yu,~\IEEEmembership{Fellow, IEEE}\vspace{-0.8cm}% <-this % stops a space
% \IEEEcompsocitemizethanks{\IEEEcompsocthanksitem 
\thanks{Lingjuan~Lyu is with Sony AI. E-mail: Lingjuan.Lv@sony.com.}
\thanks{Han Yu and Jun~Zhao are with the School of Computer Science and Engineering, Nanyang Technological University, Singapore. E-mail: han.yu@ntu.edu.sg, junzhao@ntu.edu.sg.}
\thanks{Xingjun~Ma is with the School of Information Technology, Deakin University, Geelong, Australia. E-mail: daniel.ma@deakin.edu.au.}
\thanks{Chen~Chen is with College of Computer Science, Zhejiang University, China. E-mail:cc33@zju.edu.cn.}
\thanks{Lichao~Sun is with Department of Computer Science and Engineering, Lehigh University. E-mail: lis221@lehigh.edu.}
\thanks{Qiang~Yang is with Department of AI, WeBank, Shenzhen, China, and Department of Computer Science and Engineering, Hong Kong University of Science and Technology. E-mail: qyang@ust.hk}
\thanks{Philip~S.~Yu is with Information Technology, University of Illinois at Chicago. E-mail: psyu@uic.edu.}
\thanks{$^{*}$Corresponding authors. 
}
}

\maketitle  
            
\begin{abstract}
As data are increasingly being stored in different silos and societies becoming more aware of data privacy issues, the traditional centralized training of artificial intelligence (AI) models are facing efficiency and privacy challenges. Recently, federated learning (FL) has emerged as an alternative solution and continue to thrive in this new reality. Existing FL protocol designs have been shown to be vulnerable to adversaries within or outside of the system, compromising data privacy and system robustness. Besides training powerful global models, it is of paramount importance to design FL systems that have privacy guarantees and are resistant to different types of adversaries. In this paper, we conduct a comprehensive survey on privacy and robustness in federated learning over the past 5 years. Through a concise introduction to the concept of FL, and a unique taxonomy covering: 1) threat models; 2) privacy attacks and defenses; 3) poisoning attacks and defenses, we provide an accessible review of this important topic. We highlight the intuitions, key techniques as well as fundamental assumptions adopted by various attacks and defenses. Finally, we discuss promising future research directions towards robust and privacy-preserving FL, and their interplays with multidisciplinary goals of FL.
\end{abstract}


\IEEEpeerreviewmaketitle

\section{Introduction}
\label{sec:introduction}
\IEEEPARstart{A}{s} computing devices become increasingly ubiquitous, people generate huge amounts of data through their day-to-day usage. Collecting such data into centralized storage facilities is costly and time consuming. Traditional centralized machine learning (ML) approaches cannot support such ubiquitous deployments and applications due to infrastructure shortcomings such as limited communication bandwidth, intermittent network connectivity, and strict delay constraints~\cite{li2018learning}. Another critical concern is data privacy and user confidentiality as the usage data usually contain sensitive information~\cite{abadi2016deep}. Sensitive data such as facial images, location-based services, or health information can be used for targeted social advertising and recommendation, posing immediate or potential privacy risks. Hence, private data should not be directly shared without any privacy protection. As societies become increasingly aware of privacy preservation, legal restrictions such as the General Data Protection Regulation (GDPR) are emerging, which makes data aggregation practices less feasible~\cite{FL2019}. 

In this scenario, federated learning (FL) (also well known as collaborative learning), which distributes model training to the devices from which data originate, emerged as a promising alternative ML paradigm~\cite{mcmahan2016federated}. FL enables a multitude of participants to construct a joint ML model without exposing their private training data~\cite{mcmahan2016federated,bonawitz2017practical}. It can also handle unbalanced and non-independent and identically distributed (non-I.I.D.) data, which naturally arise in the real world~\cite{mcmahan2017communication}. In recent years, FL has benefited a wide range of applications such as next word prediction~\cite{mcmahan2017communication,mcmahan2018learning}, visual object detection for safety~\cite{FedVision}, entity resolution~\cite{hardy2017private}, recommendation~\cite{wu2021fedkd,wu2020fedctr,cui2021exploiting}, industrial IoT~\cite{li2021fleam}, and graph-based analysis~\cite{wu2021fedgnn,zhou2020privacy,ni2021vertical}, etc.

\subsection{Categorization of Federated Learning based on Distribution}
Based on the distribution of data features and data samples among participants, federated learning can be generally classified as horizontally federated learning (HFL), vertically federated learning (VFL) and federated transfer learning (FTL)~\cite{yang2019federated}. 

\begin{table}[!t]
\caption{Taxonomy for horizontal federated learning (HFL).}
\label{tbl:H2B_H2C}
\centering
\scalebox{1}{
\begin{tabularx}{\linewidth}{|c|X|X|X|}
\hline
\textbf{HFL} & \textbf{Number of Participants}  & \textbf{Training Participation} & \textbf{Technical Capability} 
\tabularnewline
\hline
H2B  & small  & frequent & high
\tabularnewline
\hline
H2C & large & not frequent & low
\tabularnewline
\hline
\end{tabularx}}
\end{table}

Under HFL, datasets owned by each participant share similar features but concern different users~\cite{kantarcioglu2004privacy}. For example, several hospitals may each store similar types of data (\eg demographic, clinical, and genomics) about different patients. If they decide to build a machine learning model together using FL, we refer to such a scenario as HFL. In this paper, we further classify HFL into HFL to businesses (H2B), and HFL to consumers (H2C). A comparison between H2B and H2C is listed in Table~\ref{tbl:H2B_H2C}. The main difference lies in the number of participants, FL training participation level and technical capability, which can influence how adversaries attempt to compromise the FL system. Under H2B, there are typically a handful of participants. They can be frequently selected during FL training. The participants tend to possess significant computational power and sophisticated technical capabilities~\cite{FL2019}. Under H2C, there can be thousands or even millions of potential participants. In each round of training, only a subset of them are selected. As their datasets tend to be small, the chance of a participant being selected repeatedly for FL training is low. They generally possess limited computational power and low technical capabilities. An example of H2C is Google's GBoard application~\cite{mcmahan2018learning}.

VFL is applicable to the cases in which participants have large overlaps in the sample space but differ in the feature space, 
\ie
different participants hold different attributes of the same records~\cite{vaidya2002privacy}.
VFL mainly targets business participants. Thus, the characteristics of VFL participants are similar to those of H2B participants.

Nowadays, FTL is attracting increasing attention in industries such as finance, medicine and healthcare. FTL deals with scenarios in which FL participants have little overlap in both the sample space and the feature space~\cite{yang2019federated,FL2019}. In this case, transfer learning~\cite{pan2009survey} techniques can be applied to provide solutions for the entire sample and feature space under a federation. FTL enables complementary knowledge to be transferred across domains in a data federation, thereby enabling a target-domain party to build flexible and effective models by leveraging rich labels from a source domain~\cite{liu2020secure}.

%%%%%%%%%%%%%%%%%%%%%%%%%%%%%%%%%%%%%
\begin{figure}[!b]
\centering 
\includegraphics[width=1\columnwidth]{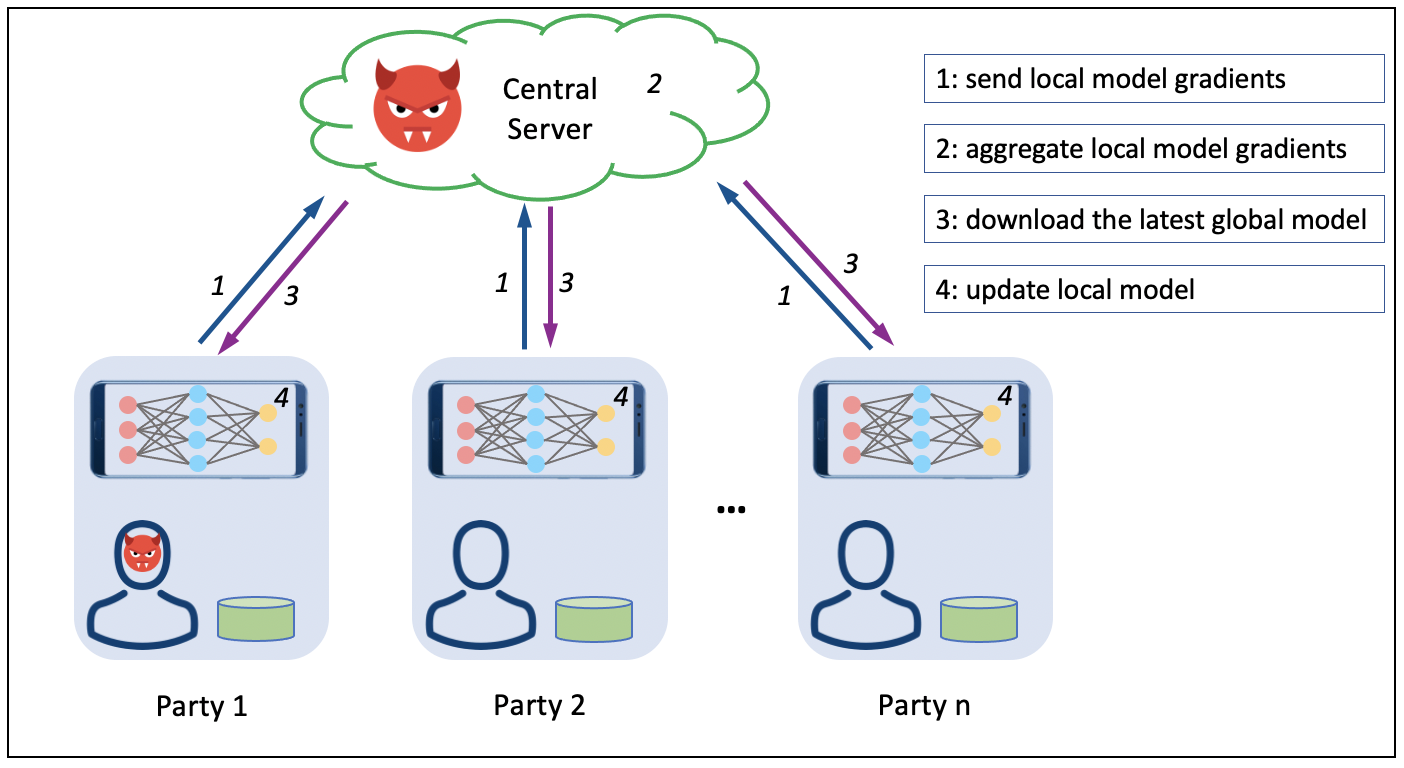}
\caption{A typical FL training process, in which both the (potentially malicious) FL server/aggregator and malicious participants may pose threats to the FL system.}
\label{fig:FL_train}
\end{figure}
%%%%%%%%%%%%%%%%%%%%%%%%%%%%%%%%%%%%%

\subsection{Categorization of Federated Learning based on Architectures}
\textbf{FL with Homogeneous Architectures:} Sharing gradients is typically limited only to homogeneous FL architectures, \ie the same model is shared with all participants. Participants aim to collaboratively learn a more accurate model. Specifically, the model parameters $w$ of the model are often obtained via solving the following optimization problem: $\min_w{\textstyle\sum}_{i=1}^n F (w, D_i)$, where $F(w, D_i)$ is the objective function for the local training dataset on the $i$-th participant and characterizes how well the parameters $w$ model the local training dataset $D_i$ on the $i$-th participant. Different classifiers (\eg logistic regression, deep neural networks) use different objective functions. In FL, each participant maintains a local model for its local training dataset. The server maintains a global model via aggregating local models from $n$ participants. Specifically, FL with homogeneous architectures performs the steps in Fig.~\ref{fig:FL_train}. FL with homogeneous architectures generally comes in two forms~\cite{mcmahan2017communication}: (1) FedSGD, in which each participant sends every SGD update to the server; (2) FedAvg, in which participants locally batch multiple iterations of SGD before sending updates to the server, which is more communication efficient. These methods are all based on the mean aggregation rule that takes the average of the local model parameters as the global model. However, the global model mean value can be arbitrarily manipulated even if just one participant is compromised~\cite{blanchard2017machine,yin2018byzantine}.

\textbf{FL with Heterogeneous Architectures:} The most recent efforts extended FL to collaboratively train models with heterogeneous architectures~\cite{Gao-et-al:2019,li2019fedmd}. Conventional federated model training which directly averages model weights is only possible if all local models have the same model structure. Naturally, it limits collaboration among data owners with heterogeneous model architectures. Sharing model prediction instead of model parameters or updates removes this obstacle and eliminates the risk of white-box inference attacks in conventional federated learning~\cite{jeong2018communication,li2019fedmd}. Unlike the existing federated learning algorithms, \emph{Federated Model Distillation} (FedMD) does not force a single global model onto local models. Instead, it is conducted in a succinct, black-box and model agnostic manner. Each local model is updated separately, participants share the knowledge of their local models via their predictions on an unlabeled public set \cite{li2019fedmd}. Another obvious benefit of sharing logits is the reduced communication costs, without significantly affecting utility~\cite{li2019fedmd}. 

In summary, all the above sharing methods did not provide defense against privacy and poisoning attacks -- two main source of threats to FL.

\subsection{Threats to FL} 
\label{sec:Threats}
FL offers a privacy-aware paradigm of model training which does not require data sharing and allows participants to join and leave a federation freely. Nevertheless, recent works have demonstrated that FL may not always provide sufficient privacy and robustness guarantees. Existing FL protocol designs are vulnerable to: (1) a malicious server who aims to infer sensitive information from individual updates over time, tamper with the training process or control the view of the participants on the global parameters; (2) any adversarial participant who can infer other participants' sensitive information, tamper the global parameter aggregation or poison the global model.

In terms of privacy leakage, communicating gradients throughout the training process can reveal sensitive information~\cite{bhowmick2018protection,melis2019exploiting}, even cause deep leakage~\cite{zhu2019deep}, either to a third party or the central server~\cite{mcmahan2018learning,agarwal2018cpsgd}. For instance, as mentioned in~\cite{aono2018privacy}, even a small portion of gradients can reveal a fair amount of sensitive information about the local data. Recent works further show that, by simply observing the gradients, a malicious attacker can successfully steal the training data~\cite{zhu2019deep,zhao2020idlg}.

In terms of robustness, FL systems are vulnerable to both data poisoning \cite{wang2020attack,xie2020} and model poisoning attacks  \cite{bagdasaryan2018backdoor,bhagoji2018analyzing,fung2020limitations,sun2019can}. 
Malicious participants can attack the convergence of the global model or implant backdoor triggers into the global model by deliberately altering their local data (data poisoning) or their gradients uploads (model poisoning). More broadly, poisoning attacks can be categorized into (1) untargeted attack such as Byzantine attack where the adversary aims to destroy the convergence and performance of the global model \cite{bernstein2019signsgd,blanchard2017machine}; and (2) targeted attack such as backdoor attack where the adversary aims to implant a backdoor trigger into the global model so as to trick the model to constantly predict an adversarial class on a subtask while keeping good performance on the main task \cite{bagdasaryan2018backdoor,bhagoji2018analyzing,xie2020}.

These privacy and robustness attacks pose significant threats to FL. In centralized learning, the server is responsible for all the participants' privacy and model robustness. However, in FL, any participant can attack the server and spy on other participants, which sometimes even without involving the server. Therefore, it is important to understand the principles behind these privacy and robustness attacks. The properties of the representative privacy and robustness attacks in server-based FL are summarized in Table \ref{tbl:attacks}.

\begin{table*}[ht]
\caption{A summary of attacks against server-based FL.}
\label{tbl:attacks}
\centering
\resizebox*{1\textwidth}{!}{
\begin{tabular}{|cc|c|c|c|c|c|c|c|c|c|}
\hline
 \multicolumn{2}{|c|}{\multirow{3}{*}{\textbf{Attack Type}}}
 &\multicolumn{2}{c|}{\textbf{Attack Target}} & \multicolumn{2}{c|}{\textbf{Attacker Role}}  & \multicolumn{2}{c|}{\textbf{FL Scenario}} & \multicolumn{3}{c|}{\textbf{Attack Complexity}}\\ 
\cline{3-11}
  & & \textbf{Model} & \textbf{Training Data} & \textbf{Participant} &\textbf{Server} & \textbf{H2B} & \textbf{H2C} & \multicolumn{2}{c|}{\textbf{Attack Iteration}} & \textbf{Auxiliary Knowledge Required} \\
\cline{9-10}
& & & & & & & &\textbf{One Round} &\textbf{Multiple Rounds} & \\
\hline
\multicolumn{1}{|c|}{\multirow{2}{*}{Robustness}} & Untargeted attack & YES & NO & YES & NO & YES & NO & YES & YES & YES \\
\cline{2-11}
\multicolumn{1}{|c|}{} & Targeted attack & YES & NO & YES & NO & YES & NO & YES & YES & YES \\
\hline
\multicolumn{1}{|c|}{\multirow{4}{*}{Privacy}} & Infer Class Representatives &NO &YES  &YES & YES & YES & NO  &NO & YES & YES \\ 
\cline{2-11}
\multicolumn{1}{|c|}{} & Infer Membership &NO &YES & YES & YES &YES & NO & NO & YES  & YES \\ 
\cline{2-11}
\multicolumn{1}{|c|}{} & Infer Properties &NO & YES & YES & YES & YES & NO &NO &YES &YES  \\ 
\cline{2-11}
\multicolumn{1}{|c|}{} & Infer Training Inputs and Labels &NO & YES & NO & YES &YES &NO &YES &YES &NO  \\ 
\hline
\end{tabular}}
\end{table*}

\subsection{Secure FL}
Attacks on FL come from either the privacy perspective when a malicious participant or the central server attempts to infer the private information of a victim participant, or the robustness perspective when a malicious participant aims to compromise the global model.

To secure FL against privacy attacks, existing privacy-preserving methodologies in centralized machine learning have been tried to be adopted into FL, including homomorphic encryption (HE), secure multiparty computation (SMC), and differential privacy (DP). However, HE and SMC may not be applicable to large-scale FL, as they incur substantial communication and computation overhead. In aggregation-based tasks, DP requires the aggregated value to contain random noise up to a certain magnitude to ensure $(\epsilon,\delta)$-DP, thus is also not ideal for FL.
The noise addition required by DP is also hard to execute in FL. In an ideal scenario where the server (aggregator) is trusted, the server can add the noise to the aggregated gradients~\cite{mcmahan2018learning}. However, in many real-world scenarios, the participants may not trust the central server nor each other. In this case, the participants would compete with each other, and all want to ensure Local Differential Privacy (LDP) by adding as much noise as possible to their local gradients~\cite{agarwal2018cpsgd,truex2018hybrid,lyu2020distributed}. This tends to accumulate significant error at the server side. \emph{Distributed Differential Privacy} (DDP)~\cite{agarwal2018cpsgd,truex2018hybrid,lyu2020distributed} can mitigate this problem to some extent when at least a certain fraction of the participants are honest and do not conduct such malicious competition.

Defending FL against various robustness attacks (\eg untargeted Byzantine attack, targeted backdoor attack) is an extremely challenging task. This is due to two main reasons. First, the defense can only be executed at the server side where only local gradients are available. This invalids many backdoor defense methods developed in the centralized machine learning, for example, denoising (preprocessing) methods \cite{liu2017neural,doan2019februus,udeshi2019model,villarreal2020confoc,li2020rethinking}, backdoor sample/trigger detection methods \cite{tran2018spectral,chen2018detecting,tang2019demon,soremekun2020exposing,chan2019poison,chou2020sentinet}, robust data augmentations \cite{liu2020reflection}, finetuning methods \cite{liu2020reflection}, neural attention distillation (NAD) based method \cite{li2021neural}, and more recent anti-backdoor learning method based on a sophisticated learning process \cite{li2021anti}. Second, the defense method has to be robust to both data poisoning and model poisoning attacks. 
Most existing robustness defenses are gradient aggregation methods mainly developed for defending against the untargeted Byzantine attackers, such as Krum/Multi-Krum \cite{blanchard2017machine}, AGGREGATHOR \cite{damaskinos2019aggregathor}, Byzantine Gradient Descent (BGD) \cite{chen2017distributed}, Median-based Gradient Descent \cite{yin2018byzantine}, Trimmed-mean-based Gradient Descent \cite{yin2018byzantine} and {\scriptsize SIGN}SGD \cite{bernstein2019signsgd}. These defense methods have never been tested on the targeted backdoor attacks \cite{bagdasaryan2018backdoor,bhagoji2018analyzing,sun2019can,wang2020attack,xie2020}.
Dedicated defense methods against both data poisoning and model poisoning attacks have been investigated, such as norm clipping \cite{sun2019can}, geometric median based Robust Federated Aggregation (RFA) \cite{pillutla2019robust} and robust learning rate \cite{ozdayi2020defending}. For the collusion of Sybil attacks, contribution similarity ~\cite{fung2020limitations} can be leveraged as a strategy for defense.

\begin{table}[!t]
\caption{A list of abbreviations used in this survey.}
\label{tbl:abbreviations}
\centering
\scalebox{1}{
\begin{tabularx}{\linewidth}{|c|X|}
\hline
AI &  Artificial Intelligence
\tabularnewline
\hline
ML &  Machine Learning
\tabularnewline
\hline
FL & Federated Learning 
\tabularnewline
\hline
GDPR & General Data Protection Regulation
\tabularnewline
\hline
I.I.D. & Independent and Identically Distributed
\tabularnewline
\hline
IoT & Internet of Things
\tabularnewline
\hline
HFL & Horizontally Federated Learning 
\tabularnewline
\hline
VFL & Vertically Federated Learning 
\tabularnewline
\hline
FTL & Federated Transfer Learning 
\tabularnewline
\hline
H2B  & HFL to businesses
\tabularnewline
\hline
H2C & HFL to consumers
\tabularnewline
\hline
SGD & Stochastic Gradient Descent
\tabularnewline
\hline
SMC & Secure Multiparty Computation
\tabularnewline
\hline
DP & Differential Privacy
\tabularnewline
\hline
CDP & Centralized Differential Privacy
\tabularnewline
\hline
LDP & Local Differential Privacy
\tabularnewline
\hline
DDP & Distributed Differential Privacy
\tabularnewline
\hline
HE & Homomorphic Encryption
\tabularnewline
\hline
RFA & Robust Federated Aggregation
\tabularnewline
\hline
GAN & Generative Adversarial Network
\tabularnewline
\hline
MIA & Membership Inference Attack
\tabularnewline
\hline
AT & Adversarial Training
\tabularnewline
\hline
FAT & Federated Adversarial Training
\tabularnewline
\hline
API & Application Programming Interface
\tabularnewline
\hline
\end{tabularx}}
\end{table}

%%%%%%%%%%%%%%%%%%%%%%%%%%%%%%%%%%%%%
\begin{figure}[!t]
\centering
\includegraphics[scale=0.38]{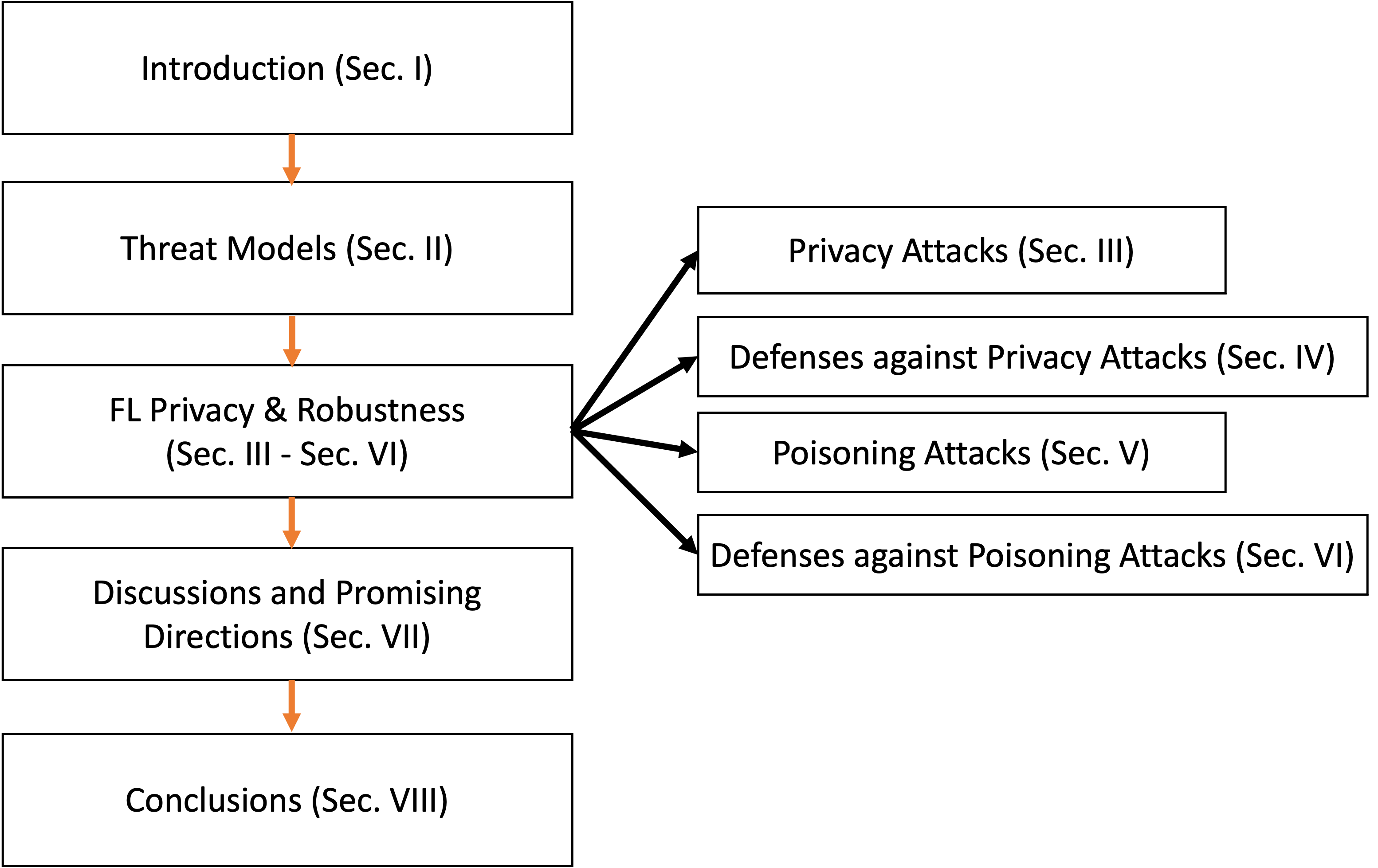}
\caption{Survey organization.}
\label{fig:flow}
\end{figure}

\subsection{Motivation of this Survey and Our Contribution}
Existing surveys on FL are mostly focused on the system or protocol design~\cite{yang2019federated,li2019federated,kairouz2019advances}. 
A notable number of research works have been conducted on privacy and robustness. Although these works attempt to discover the vulnerabilities of FL and aim to enhance the privacy and system robustness of FL, there are very few efforts for categorizing them in a systemic manner, and privacy and robustness threats to FL have not been systematically explored. To fill in this gap, in this paper, we have conducted an extensive survey on the recent advances in privacy and robustness threats to FL and their defenses. In particular, we focus on two specific threats initiated by insiders in FL systems: 1) privacy attacks that attempt to infer the victim participants' private information; 2) poisoning attacks that attempt to prevent the learning of a global model, or implant triggers to control the behaviour of the global model. This article mainly surveys the literature over the past 5 years on privacy and robustness in federated learning, it can be a notable inclusion to the existing literature, helping the community better understand the state-of-the-art privacy and robustness progress in FL. The limitations and the promising use cases of the existing works in literature, and open directions for future research are also offered to identify the research gaps to address the challenges of privacy and robustness in FL. The major contributions of this survey include:
\begin{itemize}
    \item This survey presents a comprehensive categorization of FL, and summarized threats and the corresponding protections for FL in a systematic manner.
    \item Existing privacy and robustness attacks and defenses are  well explored to help readers better understand the assumptions, principles, reasons and differences of the current progress in the domain of FL privacy and robustness. 
    \item The conflicts between privacy and robustness, and among multiple design goals are identified; the gaps between the current works and the real scenarios in FL are summarized. 
    \item Future research directions will assist community to rethink and improve their current designs towards robust and privacy-preserving FL of real practicality and impact. Meanwhile, it is suggested to integrate multidisciplinary goals in the system design of FL.
\end{itemize}

\subsection{Survey Organization}
The rest of the survey is organized as follows. Before going into an in-depth discussion on privacy and robustness in FL, in Section~\ref{sec:related_current_work}, we first summarize the threat models from a general perspective and discuss the customized threat models for privacy and robustness respectively. Section~\ref{sec:Inference} presents a comprehensive review of the privacy attacks in FL, particularly targeting the sensitive information (class representative, membership, properties, training inputs and labels) in HFL with homogeneous architectures. To address the corresponding privacy attacks, Section~\ref{sec:Privacy_Defenses} lists the most representative privacy-preserving techniques and current practices that have applied these techniques in FL. Section~\ref{sec:Poisoning} shows the detailed poisoning attacks that aim to compromise system robustness, including the untargeted and targeted poisoning attacks, followed by their countermeasures in Section~\ref{sec:Poisoning_Defenses}. From the lessons learned in this survey paper, the research gaps towards realizing trustworthy FL, along with directions for future research are provided in Section~\ref{sec:future}. Finally, concluding remarks are drawn in Section~\ref{sec:Conclusions}.

For better readability, we give a diagram in Fig.~\ref{fig:flow} showing the different aspects covered in the survey. The list of abbreviations used in this survey is provided in Table~\ref{tbl:abbreviations}. Throughout this survey, we will interchangeably use participants/clients/users to represent the participants in FL.

\section{Threat Models}
\label{sec:related_current_work}
Before reviewing attacks on FL, we first present a summary of the threat models. Generally, threat models in FL can be categorized into two types: (1) Insider v.s. Outsider; (2) Training Phase v.s. Inference Phase. These threat models apply to both the privacy and robustness. Additionally, privacy and robustness have their own threat models.

\subsection{Insider v.s. Outsider}
Attacks can be carried out by insiders and outsiders. Insider attacks include those launched by the FL server and the participants in the FL system. Outsider attacks include those launched by the eavesdroppers on the communication channel between participants and the FL server, and by users of the final FL model when it is deployed as a service.

Insider attacks are generally more dangerous than outsider attacks, as it strictly enhances the capability of the adversary. Thus, our discussion of attacks against FL will focus primarily on the insider attacks. 

\subsection{Training Phase v.s. Inference Phase}
\noindent \textbf{Training Phase.} Attacks conducted during the training phase attempt to learn, influence, or corrupt the FL model itself~\cite{biggio2011support}. 
During the training phase, the attacker can run data poisoning attacks to compromise the integrity of the training dataset~\cite{fung2020limitations,bagdasaryan2018backdoor,miao2018towards,miao2018attack,zhang2019data,sun2021data}, or model poisoning attacks to compromise the integrity of the learning process~\cite{bhagoji2018analyzing,fang2020local}. The attacker can also launch a range of inference attacks on an individual participant's update or on the aggregated update from all participants during training phase~\cite{melis2019exploiting}.

\noindent \textbf{Inference Phase.} Attacks conducted during the inference phase are called evasion or exploratory attacks~\cite{barreno2006can}. They generally do not alter the targeted model, instead, either trick it to produce wrong predictions (targeted/untargeted) or collect evidence about the model characteristics, causing privacy and robustness problems. The effectiveness of such attacks are largely determined by the information that is available to the adversary about the model.
Inference phase attacks can be classified into white-box attacks (\ie with full access to the FL model) and black-box attacks (\ie only able to query the FL model). In FL, the global model maintained by the server suffers from the same evasion attacks as in the conventional ML setting when the target model is deployed as a service. While black-box attacks may be more natural to consider in the centralized settings, the model broadcast step in FL makes the global model a white-box to any malicious participant. Thus, FL requires extra efforts to defend against white-box evasion attacks~\cite{kairouz2019advances}.

\subsection{Privacy: Semi-honest v.s. Malicious} 
\noindent \textbf{Semi-honest Setting.} Adversaries are considered passive or honest-but-curious. They try to learn the private states of other participants without deviating from the FL protocol. The adversaries can only observe the received information, \ie parameters of the global model. 

\noindent \textbf{Malicious Setting.} An active or malicious adversary tries to learn the private states of honest participants, and deviates arbitrarily from the FL protocol by modifying, re-playing, or removing messages. This setting allows the adversary to conduct particularly devastating attacks.

\subsection{Robustness: Untargeted v.s. Targeted} 
\begin{enumerate}
    \item Untargeted attack: The untargeted poisoning attack aims to arbitrarily compromise the integrity of the target model. Byzantine attack is one type of the untargeted poisoning attacks that uploads arbitrarily malicious gradients to the server so as to cause the failure of the global model~\cite{lamport1982byzantine,blanchard2017machine,damaskinos2019aggregathor,bernstein2019signsgd,xie2020fall}. 
    \item Targeted attack: The targeted poisoning attack induces the model to output the target label specified by the adversary for particular testing examples, while the testing error for other testing examples is unaffected~\cite{bagdasaryan2018backdoor,bhagoji2018analyzing,fung2020limitations,sun2019can}. 
\end{enumerate}

\section{Privacy Attacks}
\label{sec:Inference}
\begin{figure}[!t]
\centering
\includegraphics[scale=0.38]{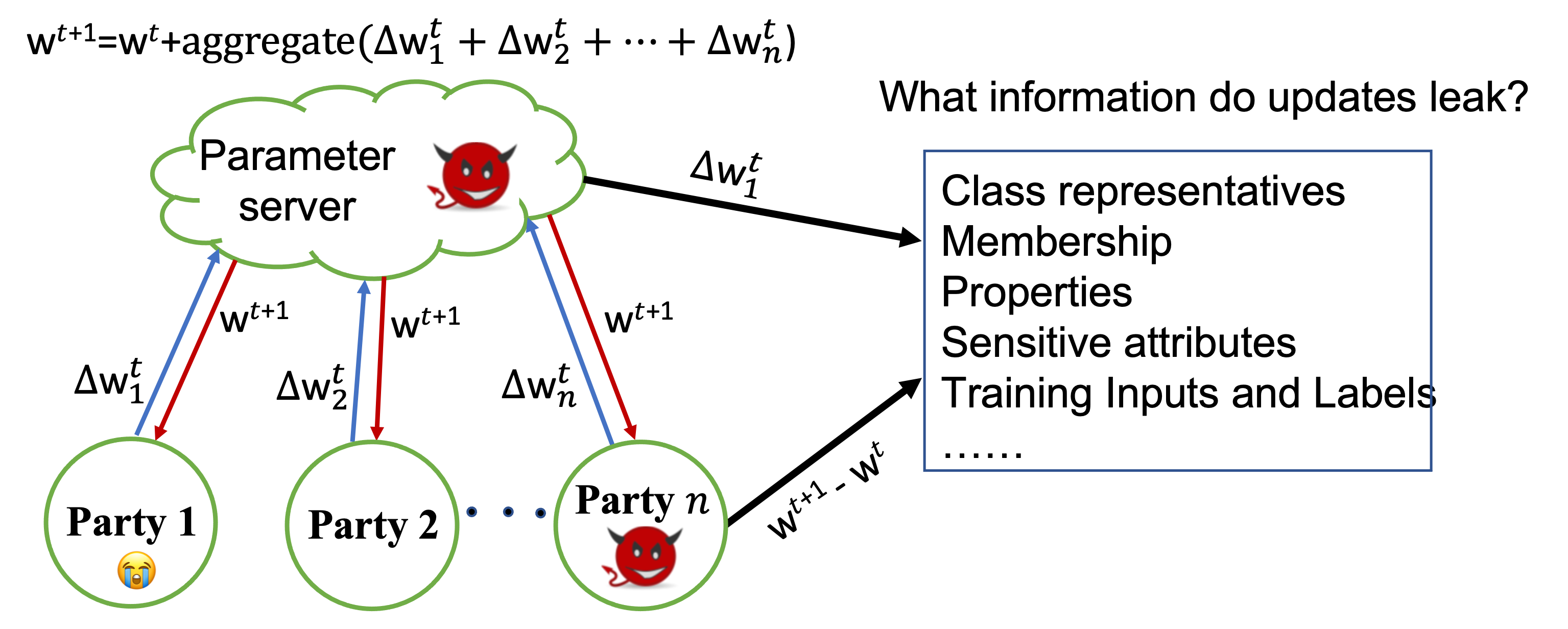}
\caption{A demo of privacy leakage in FL. Attacker can infer various private information about the victim participant from the received gradients or the snapshot of the FL model parameters.}
\label{fig:inference_FL}
\end{figure}

Although FL prevents the participants from directly sharing their private data, a series of works have demonstrated that exchanging gradients in FL can also leak sensitive information about the participants’ private data to either passive or active attackers~\cite{phong2018privacy,su2018securing,melis2019exploiting,zhu2019deep,nasr2019comprehensive}. For example, gradients or two consecutive snapshots of the FL model parameters can leak unintended features of the participants' training data to the adversarial participants, as deep learning models tend to recognize and remember more features of the data than needed for the main learning task~\cite{zhang2017understanding}. Fig.~\ref{fig:inference_FL} illustrates the set of information an adversary can infer from the gradients (\ie $\Delta \boldsymbol{w}_1^t$), or equivalently the difference of two successive snapshots of the model parameters (\ie $\boldsymbol{w}^{t+1}-\boldsymbol{w}^t$).
%%%%%%%%%%%%%%%%%%%%%%%%%%%%%%%%%%%%%
%%%%%%%%%%%%%%%%%%%%%%%%%%%%%%%%%%%%%

The reason why gradients can cause privacy leakage is that the gradients are derived from the participants' private training data, and a learning model can be considered as a representation of the high-level statistics of the dataset it was trained on~\cite{lyu2018privacy_thesis}.
In deep learning models, gradients of a given layer are computed based on the layer's features and the error from the layer after (\ie backpropagation). In the case of sequential fully-connected layers, the gradients of the weights are the inner products of the current layer's features and the error from the layer after. Similarly, for a convolutional layer, the gradients of the weights are convolutions of the layer's features and the error from the layer after~\cite{melis2019exploiting}. Consequently, observations of gradients can be used to infer a significant amount of private information, such as class representatives, membership and properties of a subset of the training data. Even worse, an attacker can infer labels from the shared gradients and recover the original training samples without any prior knowledge about the training data~\cite{zhu2019deep}. Next, we detail the potential privacy leakage of FL according to the type of the sensitive information that the attacker is targeting at.

\subsection{Inferring Class Representatives}
Hitaj \etal~\cite{hitaj2017deep} firstly devised an active inference attack called \emph{Generative Adversarial Networks} (GAN) attack against deep FL models. In this attack, a malicious participant can intentionally compromise any other participant. 
The GAN attack exploits the real-time nature of the FL learning process which allows the adversarial participant to train a GAN to generate prototypical samples of the targeted private training data. The generated samples appear to come from the same distribution as the training data. Hence, GAN attack is not targeted to reconstruct the exact training inputs, but only the class representatives. It should be noted that GAN attack assumes the entire training corpus for a given class comes from a single participant, which means the GAN-constructed representatives are similar to the training data only when all class members are similar. This resembles model inversion attacks in the centralized ML settings~\cite{fredrikson2015model}. Note that these assumptions may be less practical in FL. Since GAN attack requires a substantial amount of computational resources to train the GAN model, it is less suitable for H2C scenarios.

\subsection{Inferring Membership}
Given an exact data point, membership inference attacks (MIA) aim to determine if it was used to train the model~\cite{shokri2017membership}. For example, an attacker can infer whether a specific patient profile was used to train a classifier associated with a certain disease. FL opens new possibilities for such attacks. In FL, the adversary can infer if a particular sample belongs to the private training data of a particular participant (if the target update is from a single participant) or any participant (if the target update is the aggregate). For example,  during FL model training, the non-zero gradients of the embedding layer of a deep natural language processing model trained on text data can reveal which words are in the training batches of the honest participants~\cite{melis2019exploiting}. 

Attackers in a FL system can conduct both active and passive membership inference attacks~\cite{nasr2019comprehensive,melis2019exploiting}. In the passive case, the attacker observes the updated model parameters and performs inference without modifying the learning process. In the active case, the attacker can tamper with the FL model training protocol and perform a more powerful attack against other participants. For instance, the attacker may share malicious updates and trick the FL model to expose more information about other participants' local data. One such attack is the gradient ascent attack~\cite{nasr2019comprehensive}, where the attacker runs gradient ascent on a target data sample and observes whether its increased loss can be drastically reduced in the next communication round, if so, the sample is very likely to be in the training set. This attack can be applied on a batch of target data samples all at the same time \cite{nasr2019comprehensive}.

\subsection{Inferring Properties} 
An adversary can launch both passive and active property inference attacks to infer certain properties of other participants' training data~\cite{melis2019exploiting}. Property inference attacks assume that the adversary has auxiliary training data that are correctly labelled with the target property. A passive adversary can only observe or eavesdrop the gradients and perform inference by training a binary property classifier. An active adversary can exploit multi-task learning to trick the FL model into learning a better separation between data with and without the target property so as to extract more information. An adversarial participant can also infer when a property appears or disappears in the training data (\eg identifying when a person first appears in the photos used to train a gender classifier). The assumption of auxiliary training data in property inference attacks may limit its applicability in H2C. %\ma{what assumption?}

\subsection{Inferring Training Inputs and Labels}
One recent work called \emph{Deep Leakage from Gradient} (DLG) proposes an optimization algorithm to extract both the training inputs and the labels~\cite{zhu2019deep}. This attack is much stronger than previous approaches. It can accurately recover the raw images and texts used to train a deep learning model. In a follow-up work \cite{zhao2020idlg}, an analytical approach called \emph{Improved Deep Leakage from Gradient} (iDLG) was proposed to extract labels based on the shared gradients and an exploration of the correlation between the labels and the signs of the gradients. iDLG can be applied to attack any differentiable models trained with cross-entropy loss and one-hot labels, which is a typical setting for classification tasks.

In summary, inference attacks generally assume that the adversaries possess sophisticated technical capabilities and unlimited computational resources. Moreover, most attacks assume that the adversarial participants can be selected (to update the global model) in many rounds of the FL training process. In FL, these assumptions are generally not practical in H2C scenarios, but more likely happen in H2B scenarios. These inference attacks highlight the need for gradient protection in FL, possibly through various privacy-preserving mechanisms~\cite{FL2019} detailed in Section~\ref{sec:Privacy_Defenses}.

\section{Defenses against Privacy Attacks}
\label{sec:Privacy_Defenses}

 \begin{table}[ht]
 \caption{Privacy-preserving Techniques for FL.}
 \label{tbl:pp_techniques}
 \centering
 \begin{tabular}{|c|c|c|}
 \hline
 \multicolumn{2}{|c|}{\textbf{Privacy-preserving Techniques}} & \textbf{Existing Works}
 \\
 \hline
 \multicolumn{2}{|c|}{Homomorphic Encryption} &  ~\cite{aono2016scalable,kim2018secure} 
  \\ 
 \hline
\multirow{3}{*}{\makecell{DP}} & CDP &~\cite{geyer2017differentially,mcmahan2018learning}
\tabularnewline
\cline{2-3}
& LDP & ~\cite{nguyen2016collecting,wang2019collecting,zhao2020local,bhowmick2018protection,sun2020ldp,truex2020ldp,sun2020federated}
\tabularnewline
\cline{2-3}
& DDP+Cryptography & ~\cite{agarwal2018cpsgd,truex2018hybrid,lyu2020lightweight}
  \\
 \hline
 \multicolumn{2}{|c|}{Secure Multiparty Computation} &
 ~\cite{mohassel2017secureml,bonawitz2017practical} 
 \\ 
 \hline
 \end{tabular}
 \end{table}
 
While privacy preservation has been extensively studied in the machine learning community, privacy preservation in federated learning can be more challenging due to the sporadic access to power and network connectivity, statistical heterogeneity in the data, etc. Existing works in privacy-preserving federated learning are mostly developed based on the well-known privacy-preserving techniques, including: (1) \emph{homomorphic encryption} (HE), such as Paillier~\cite{paillier1999public}, Elgamal~\cite{elgamal1985public} and Brakerski-Gentry-Vaikuntanathan cryptosystems~\cite{gentry2009fully}; (2) \emph{Secure Multiparty Computation} (SMC), such as garbled circuits~\cite{yao1982protocols} and secret sharing~\cite{demmler2015aby}; and (3) \emph{differential privacy} (DP)~\cite{dwork2006calibrating,dwork2014algorithmic}. A concise summary of privacy-preserving techniques is listed in Table~\ref{tbl:pp_techniques}.

\subsection{Privacy Preservation through Homomorphic Encryption} 
A homomorphic encryption scheme allows arithmetic operations to be directly performed on ciphertexts, which is equivalent to a specific linear algebraic manipulation of the plaintext. 
Existing homomorphic encryption techniques can be categorized into: 1) fully homomorphic encryption, 2) somewhat homomorphic encryption, and 3) partially homomorphic encryption. Fully homomorphic encryption can support arbitrary computation on ciphertexts, but is less efficient~\cite{gentry2009fully}. On the other hand, somewhat homomorphic encryption and partially homomorphic encryption are more efficient but are specified by a limited number of operations~\cite{damgaard2012multiparty,rivest1978method,elgamal1985public,paillier1999public}. Partially homomorphic encryption schemes are more widely used in practice, including RSA~\cite{rivest1978method}, El Gamal~\cite{elgamal1985public}, Paillier~\cite{paillier1999public}, etc. 
The homomorphic properties can be described as:

 \begin{equation*}\label{eq:add_hor}
 E_{pk}(m_1+m_2) = c_1 \oplus c_2 \\
 \end{equation*}
 \begin{equation*}\label{eq:mul_hor}
 E_{pk}(a \cdot m_1) = a \otimes c_1
 \end{equation*}
 where $a$ is a constant, $m_1$, $m_2$ are the plaintexts that need to be encrypted, $c_1$, $c_2$ are the ciphertext of $m_1$, $m_2$ respectively.

Homomorphic encryption is widely used and is especially useful for securing the learning process by computing on encrypted data. However, doing arithmetic on the encrypted numbers comes at a cost of memory and processing time. For example, with Paillier encryption scheme, the encryption of an encoded floating-point number (whether single or double precision) is $2m$ bits long, where $m$ is typically at least 1024 and the addition of two encrypted numbers is 2$\sim$3 orders of magnitude slower than the unencrypted equivalent~\cite{hardy2017private}.
Moreover, polynomial approximations need to be made to evaluate non-linear functions in machine learning algorithms, resulting in a trade-off between utility and privacy~\cite{aono2016scalable,kim2018secure}. For example, to protect individual gradients, Aono \etal~\cite{aono2018privacy} used additively homomorphic encryption to preserve the privacy of gradients and enhance the security of the distributed learning system. However, their protocol not only incurs large communication and computational overhead, but also results in utility loss. Furthermore, it is not able to withstand collusion between the server and multiple participants. Hardy \etal~\cite{hardy2017private} applied federated logistic regression on vertically partitioned data encrypted with an additively homomorphic scheme to secure against an honest-but-curious adversary. Overall, all these works incur extra communication and computational overheads, which limit their applications in H2C scenarios.
 
\subsection{Privacy-Preservation through SMC} 
\emph{Secure Multiparty Computation} (SMC)~\cite{yao1982protocols} enables different participants with private inputs to perform a joint computation on their inputs without revealing them to each other. Mohassel \etal~\cite{mohassel2017secureml} proposed SecureML which conducts privacy-preserving learning via SMC, where data owners need to process, encrypt and/or secret-share their data among two non-colluding servers in the initial setup phase. SecureML allows data owners to train various models on their joint data without revealing any information beyond the outcome. However, this comes at a cost of high computation and communication overhead, which may hamper participants' interest to collaborate. Bonawitz \etal~\cite{bonawitz2017practical} proposed a secure, communication-efficient, and failure-robust protocol based on SMC for secure aggregation of individual gradients. It ensures that the only information about the individual users the server learns is what can be inferred from the aggregated results. The security of their protocol is maintained under both the honest-but-curious and malicious settings, even when the server and a subset of users act maliciously -- colluding and deviating arbitrarily from the protocol. That is, no party learns anything more than the sum of the inputs of a subset of honest users of a large size~\cite{bonawitz2017practical}. 
 
In general, SMC techniques ensure a high level of privacy and accuracy, at the expense of high computation and communication overhead, thereby doing a disservice to attracting participation. Another main challenge facing SMC-based schemes is the requirement for simultaneous coordination of all participants during the entire training process. Such a multi-party interaction model may not be desirable in practical settings, especially under the commonly considered participant-server architecture in FL settings. Besides, SMC-based protocols can enable multiple participants to collaboratively compute an agreed-upon function without leaking input information from any participant except for what can be inferred from the outcomes of the computation~\cite{goryczka2015comprehensive,riazi2018chameleon}. That said, SMC cannot fully guarantee protection from information leakage, which requires additional differential privacy techniques to be incorporated into the multi-party protocol to address such concerns~\cite{rastogi2010differentially,shi2011privacy,acs2011have,lyu2018ppfa}.

In summary, homomorphic encryption or SMC-based approaches may not be applicable to large-scale FL scenarios as they incur substantial additional communication and computation costs. Moreover, encryption based techniques need to be carefully designed and implemented for each operation in the target learning algorithm~\cite{chen2019secure,mohassel2018aby}. Lastly, all the cryptography based protocols prevent anyone from auditing participants' updates to the joint model, which leaves spaces for
the malicious participants to attack. For example, malicious participants can introduce stealthy backdoor functionality into the global model without being detected~\cite{bhagoji2018analyzing}.

\subsection{Privacy-Preservation through Differential Privacy}
Differential privacy (DP) was originally designed for the single database scenario, where for every query made, a database server answers the query in a privacy-preserving manner with tailored randomization~\cite{dwork2006calibrating}. In comparison with encryption based approaches, differential privacy trades off privacy and accuracy by perturbing the data in a way that (i) is computationally efficient, (ii) does not allow an attacker to recover the original data, and (iii) does not severely affect the utility.

The concept of differential privacy is that the effect of the presence or the absence of a single record on the output likelihood is bounded by a small factor $\epsilon$. As defined in Definition~\ref{def:dp}, $(\epsilon,\delta)$-approximate differential privacy~\cite{dwork2014algorithmic} relaxes pure $\epsilon$-differential privacy by a $\delta$ additive term, which means the unlikely responses need not satisfy the pure differential privacy criterion. 

 \begin{definition}\label{def:dp}
 $(\epsilon,\delta)$-differential privacy~\cite{dwork2014algorithmic}. For scalars $\epsilon>0$ and $0\leq \delta<1$, mechanism \mech is said to preserve (approximate) $(\epsilon,\delta)$-differential privacy if for all adjacent datasets $D,D'\in\mathcal{D}^n$ and measurable $S \in \mathrm{range}(\mech)$,
 \begin{eqnarray*}
 \Pr\{\mech(D)\in S\} &\leq& \exp(\epsilon) \cdot \Pr\{\mech(D')\in S\} + \delta\enspace.
 \end{eqnarray*}
 \end{definition}

To avoid the worst-case scenario of always violating privacy of a $\delta$ fraction, the standard recommendation is to choose $\delta \ll 1/|D|$, where $|D|$ is the size of the database. This strategy forecloses possibility of one particularly devastating outcome, but other forms of information leakage remain.

The privacy community generally categorizes DP into the following three categories as per different trust assumptions and noise sources: centralized DP (CDP), local DP (LDP) and distributed DP (DDP). A comprehensive comparison among CDP, LDP and DDP is listed in Table~\ref{tbl:DP_Comparison}.

 \begin{table*}[ht]
 \renewcommand{\arraystretch}{1.1}
 \caption{Comparative analysis among CDP, LDP and DDP.}
 \label{tbl:DP_Comparison}
 \centering
 \scalebox{0.9}{
 \begin{tabular}{|c|c|c|c|c|}
 \hline
 \textbf{DP type} & \textbf{Trusted aggregator?} 
 & \textbf{Who should add noise?} & \textbf{Privacy Guarantee} &\textbf{Error Bound}
 \\
 \hline
 CDP~\cite{geyer2017differentially,mcmahan2018learning} & Yes   
 & aggregator & aggregated value & $O(\frac{1}{\epsilon})$ 
  \\ 
 \hline
 LDP~\cite{bhowmick2018protection,li2019differentially} & No 
 & user & locally released value & $O(\frac{\sqrt{n}}{\epsilon})$
  \\
 \hline
 DDP~\cite{agarwal2018cpsgd,lyu2020distributed} & No 
 & user & aggregated value & $O(\frac{1}{\epsilon})$  
 \\ 
 \hline
 \end{tabular}}
 %\end{tabularx}
 \end{table*}

\textbf{Centralized Differential Privacy (CDP)}. CDP was originally designed for the centralized scenario where a trusted database server, who is entitled to see all participants' data in the clear, wishes to \emph{answer queries or publish statistics} in a privacy-preserving manner by randomizing query results~\cite{dwork2006calibrating,papernot2017semi,papernot2018scalable}. When CDP meets FL, CDP assumes a trusted aggregator, who is responsible for adding noise to the aggregated local gradients to ensure record-level privacy of the whole data of all participants~\cite{mcmahan2018learning,geyer2017differentially}.
However, CDP is geared to tackle thousands of users for training to converge and achieve an acceptable trade-off between privacy and accuracy~\cite{mcmahan2018learning}, resulting in a convergence problem with a small number of participants~\cite{lyu2020threats}. Moreover, CDP can achieve acceptable accuracy only with a large number of participants, thus not applicable to H2B with relatively a small number of participants. 

Meanwhile, the assumption of a trusted server in CDP is ill-suited in many applications as it constitutes a single point of failure for data breaches, and saddles the trusted curator with legal and ethical obligations to keep the user data secure. When the aggregator is untrusted which is often the case in distributed scenarios, Local Differential Privacy (LDP)~\cite{duchi2013local} or \emph{Distributed Differential Privacy} (DDP) are needed~\cite{dwork2006our,shi2011privacy} to protect privacy of individuals. 

\textbf{Local Differential Privacy (LDP)}. Local differential privacy (LDP)~\cite{duchi2013local} offers stronger privacy guarantee, data owners perturb their private information to satisfy DP locally before reporting it to an untrusted data curator~\cite{sun2020ldp,lyu2020differentially,truex2020ldp}. A comprehensive survey of LDP can be referred to~\cite{yang2020local}. A formal definition of LDP is given in Definition~\ref{def:ldp}.

 \begin{definition}
  \label{def:ldp}
 ($(\epsilon,\delta)$-Local Differential Privacy). A randomized algorithm \mech satisfies $(\epsilon,\delta)$-local differential privacy ($(\epsilon,\delta)$-LDP) if and only if for any input $v$ and $v'$, we have
 \begin{eqnarray*}
 \Pr\{\mech(v)=o\} &\leq& \exp(\epsilon) \cdot \Pr\{\mech(v')=o\} +
 \delta\enspace
 \end{eqnarray*}

 for $\forall o \in Range(\mech)$, where $Range(\mech)$ denotes the set of all possible outputs of the algorithm \mech. Furthermore \mech is said to preserve (pure) $\epsilon$-LDP if the condition holds for $\delta=0$.
 \end{definition}

Although the randomized response~\cite{warner1965randomized} and its variants~\cite{erlingsson2014rappor} have been widely used to provide LDP when individuals disclose their personal information. We remark that all the randomization mechanisms used for CDP, such as Laplace mechanism and Gaussian mechanism~\cite{dwork2014algorithmic}, can be individually used by each participant to  ensure LDP in isolation. However, in distributed scenario, without the help of cryptographic techniques, each participant has to add enough calibrated noise to ensure LDP.
The attractive privacy properties of LDP thus come with a huge utility degradation, especially with billions of individuals. Under the CDP model, the aggregator releases the aggregated value with an expected additive error of at most $\Theta(1/\epsilon)$ to ensure $\epsilon$-DP (\eg using the Laplace mechanism~\cite{dwork2014algorithmic}). In contrast, under the LDP model, at least $\Omega(\sqrt{n}/\epsilon)$ additive error in expectation must be incurred by any $\epsilon$-DP mechanism for the same task~\cite{chan2012optimal,duchi2013local}. This gap is essential for eliminating the trust in the centralized server, and cannot be removed by algorithmic improvement~\cite{chan2019foundations}.

Several prior works have attempted to apply LDP to FL.
For example, Shokri et al. ~\cite{shokri2015privacy} firstly applied LDP to distributed/federated learning, in which each participant individually adds noise to its gradients before releasing to the server, thus ensuring local DP. However, their privacy bounds are given per-parameter, the large number of parameters prevents their method from providing a meaningful privacy guarantee~\cite{papernot2017semi}. Other approaches that also considered to apply LDP to FL can only support shallow models such as logistic regression and only focus on simple tasks and datasets~\cite{nguyen2016collecting,wang2019collecting,zhao2020local}. Bhowmick et al. ~\cite{bhowmick2018protection} presented a viable approach to large-scale local private model training, and introduced a relaxed version of LDP by limiting the power of potential adversaries. Due to the high variance of their mechanism, it requires more than 200 communication rounds and incurs much higher privacy cost, \ie MNIST ($\epsilon = 500$) and CIFAR-10 ($\epsilon = 5000$). Note that the $\epsilon$ required in~\cite{bhowmick2018protection} is relatively large, as they considered only privacy protection against \emph{reconstruction attacks} instead of membership attacks, and they pointed out that local privacy as traditionally employed may prove too stringent for practical use, especially in modern high-dimensional statistical and machine learning problems. Their obtained results suggested that using LDP mechanisms with \emph{large} $\epsilon$ may still provide decent protection against reconstruction. Li \etal~\cite{li2019differentially} proposed locally differentially-private algorithms in the context of meta-learning, which might be applicable to FL with personalization. However, it only provides provable learning guarantees in convex settings. Truex \etal~\cite{truex2020ldp} applied Condensed Local Differential Privacy ($\alpha$-CLDP) into FL. However, $\alpha$-CLDP required privacy budget $\epsilon = \alpha \cdot 2c \cdot 10^p$ (e.g. $\alpha = 1, c = 1, p = 10$), which results in a weak privacy guarantee. Another contemporary work called LDP-FL~\cite{sun2020ldp} achieves better performance on both effectiveness and efficiency than~\cite{truex2020ldp} with a special communication design for deep learning approaches. 

In addition to the privacy leakage in FL with homogeneous architectures, FL with heterogeneous architectures suffers from the similar privacy issues. In FL with homogeneous architectures, the predictions from local models are also sensitive and may leak private information~\cite{papernot2017semi,lyu2018privacy_thesis,lyu2020distributed,sun2020federated}. Currently, there is no theoretic guarantee that sharing prediction is private and secure~\cite{lyu2018privacy_thesis,sun2020federated}. To address this issue, one naive approach is adding the locally differentially private random noise to the predictions like previous works. Although the privacy concern is mitigated with random noise perturbation, it brings a new problem with a substantial trade-off between privacy budget and model utility. Sun \etal~\cite{sun2020federated} filled in this gap by proposing a novel framework called FEDMD-NFDP, which integrated a novel Noise-Free Differential Privacy (NFDP) mechanism into federated model distillation. The LDP guarantee of NFDP roots in local data sampling process, which explicitly eliminates noise addition and privacy cost explosion issues in previous works.

 %%%%%%%%%%%%%%%%%%%%%%%%%%%%%%%%%%%%%
 \begin{figure*}[!htp]
 \centering
         \begin{subfigure}{0.45\textwidth}
                \centering \includegraphics[scale=0.4]{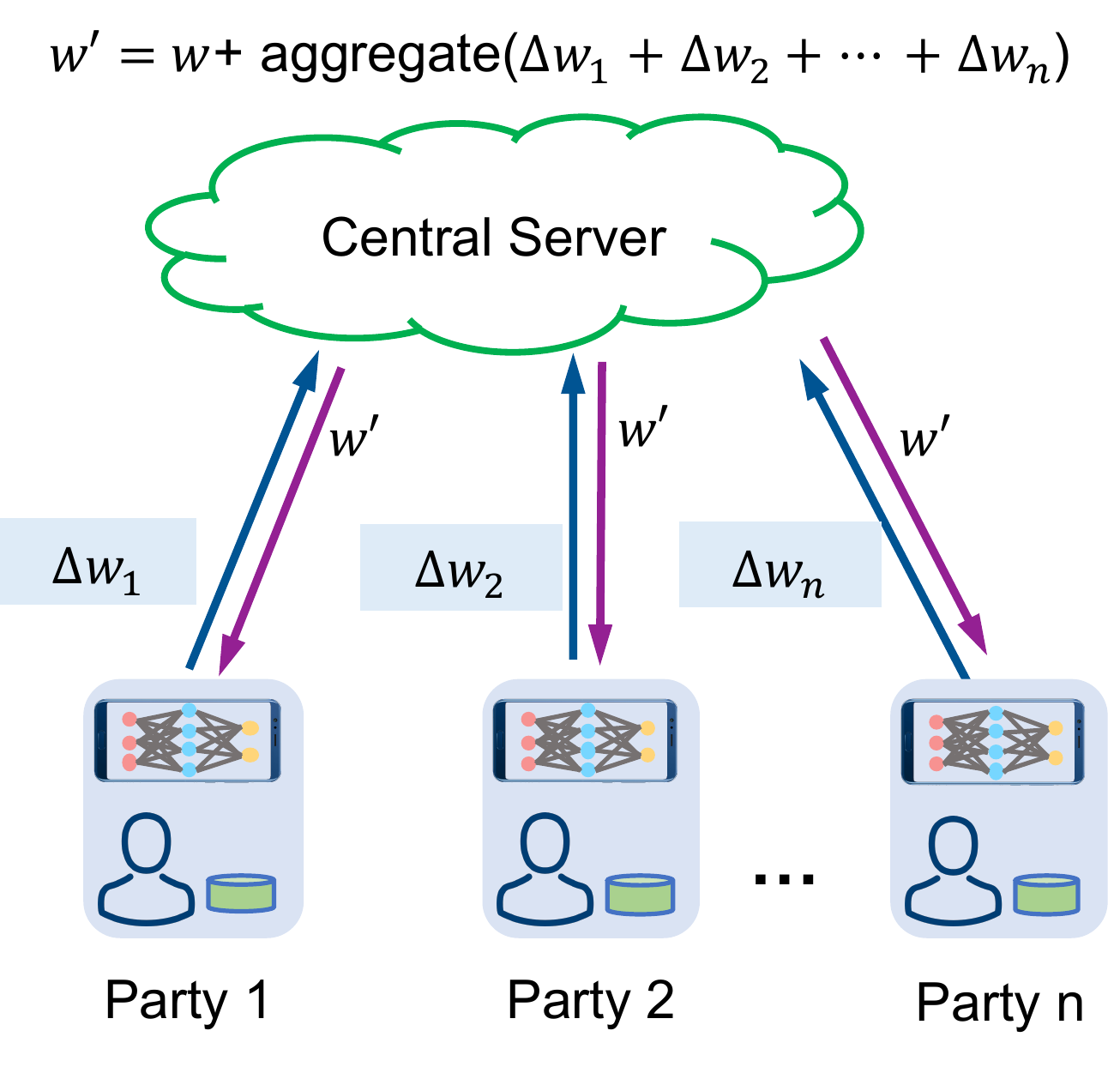}
                 \caption{FL without privacy.}
                 \label{fig:FL_noDP}
         \end{subfigure}
        %  ~~~~
          \begin{subfigure}{0.45\textwidth}
                \centering \includegraphics[scale=0.4]{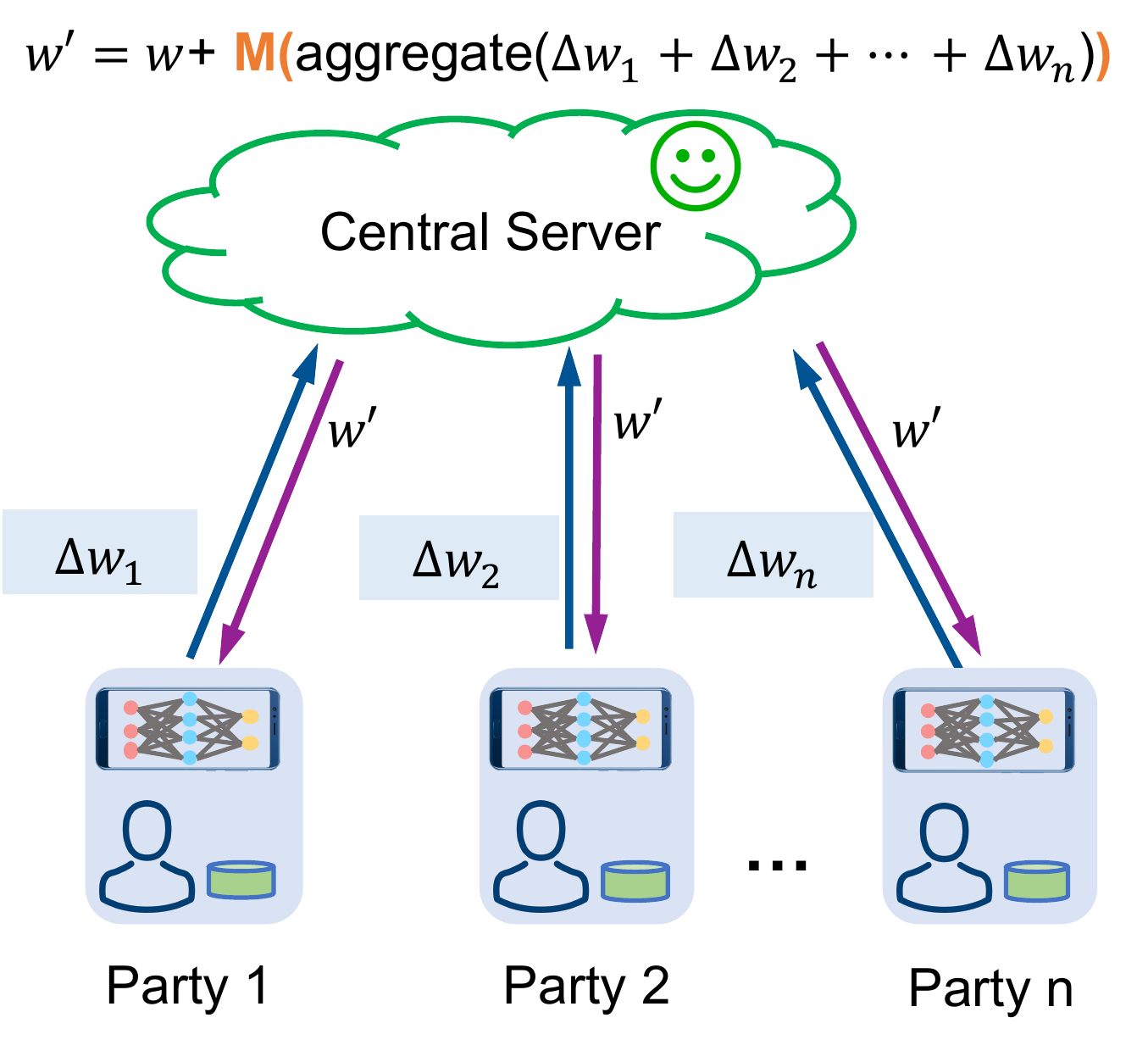}
                 \caption{Centralized DP: FL with a trusted server.}
                 \label{fig:FL_globalDP}
         \end{subfigure}
        %  ~~~~
          \begin{subfigure}{0.45\textwidth}
                \centering \includegraphics[scale=0.4]{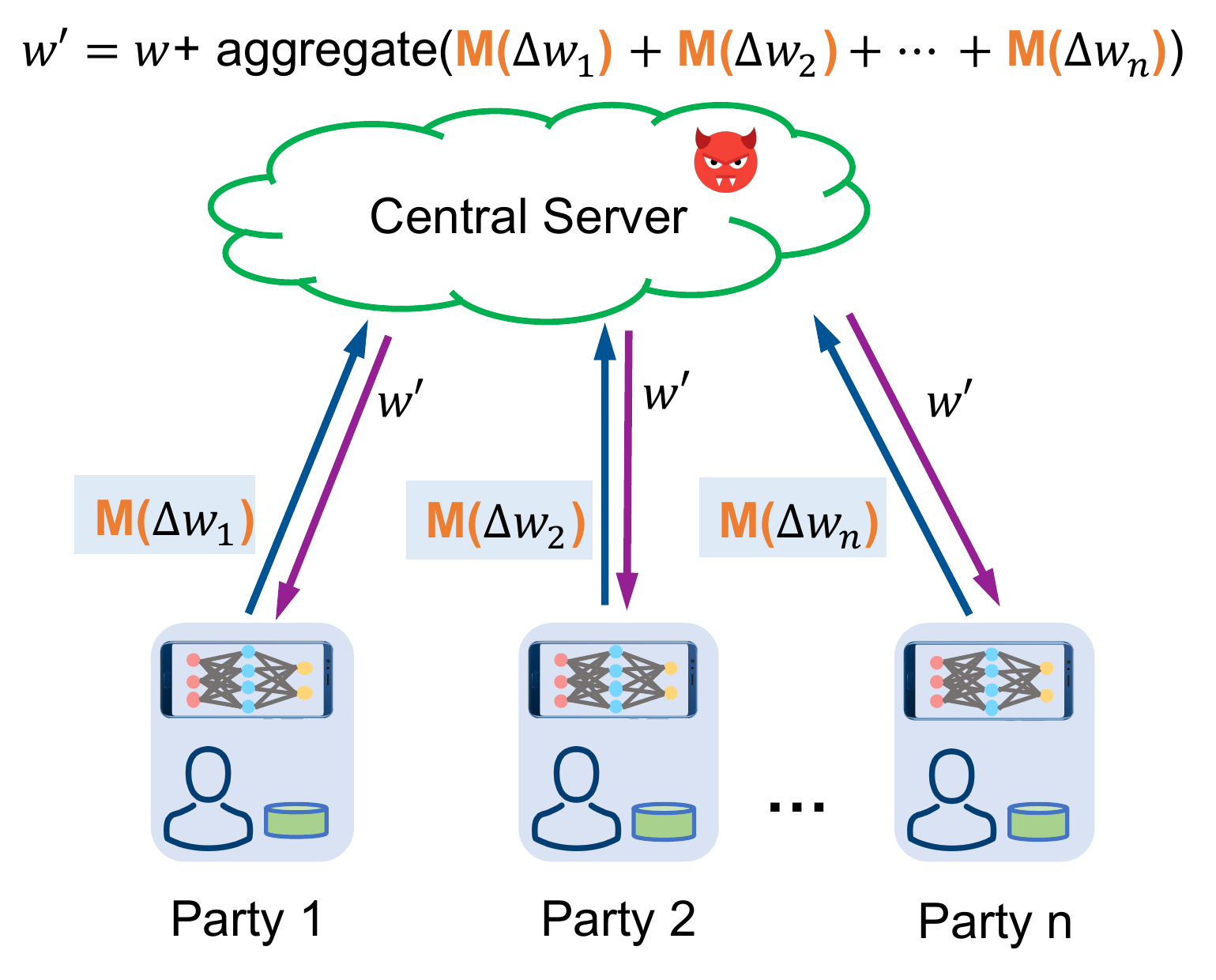}
                 \caption{Local DP: FL without a trusted server.}
                 \label{fig:FL_localDP}
         \end{subfigure}
        %  ~~~~
          \begin{subfigure}{0.45\textwidth}
                \centering \includegraphics[scale=0.4]{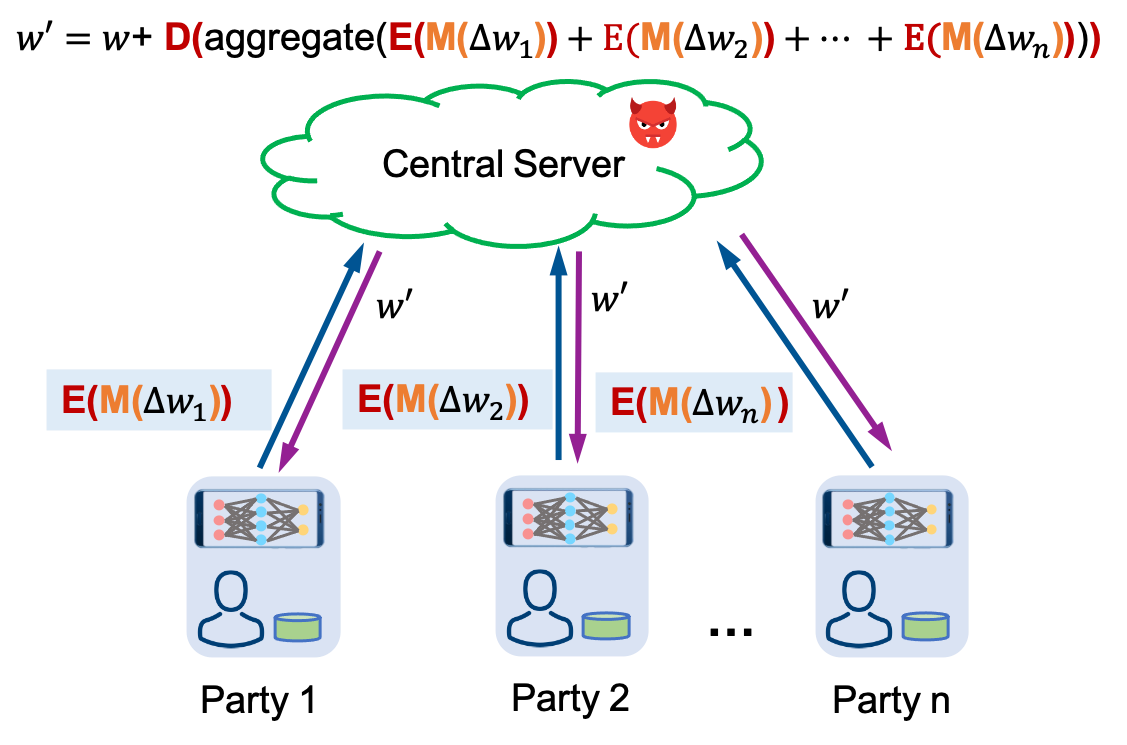}
                 \caption{Distributed DP with SMC: FL without a trusted server.}
                 \label{fig:FL_DDP}
         \end{subfigure}
         \caption{An illustration of one round of federated learning without privacy and with different DP mechanisms. $M$ denotes a DP mechanism used to privatize the data. In centralized DP (b), the central server is trusted. In local DP (c), the central server is not trusted, gradients are perturbed to ensure local DP before forwarding to the central server. In distributed DP (d),  the central server is also not trusted, gradients are perturbed via DP mechanism $M$ and encrypted via encryption operation $E$ to ensure privacy before forwarding to the central server, who needs to finally decrypt ($D$) the aggregated ciphertext.}
 \label{fig:FL_DP}
 \end{figure*}
 %%%%%%%%%%%%%%%%%%%%%%%%%%%%%%%%%%%%%
 
\textbf{Distributed Differential Privacy (DDP).}  
\emph{Distributed Differential Privacy} (DDP) bridges the gap between LDP and CDP, while ensuring the privacy of each individual by combining with cryptographic protocols~\cite{rastogi2010differentially,shi2011privacy,acs2011have,lyu2018ppfa,agarwal2018cpsgd}. Therefore, DDP avoids placing trust in any server, and offers better utility than LDP. Theoretically, DDP offers the same utility as CDP, as the total amount of noise is the same.

The notion of DDP reflects the fact that the required noise in the target statistic is sourced from multiple participants~\cite{dwork2006our}. 
Approaches to DDP that implement an overall additive noise mechanism by summing the same mechanism run at each participant (typically with less noise) necessitate mechanisms with stable distributions---to guarantee proper calibration of known end-to-end response distribution---and cryptography for hiding all but the final result from participants~\cite{shi2011privacy,dwork2006our,acs2011have,rastogi2010differentially,agarwal2018cpsgd,truex2018hybrid,lyu2020lightweight}.
Stable distributions include Gaussian distribution, Binomial distribution~\cite{lyu2020distributed}, etc, \ie sum of Gaussian random variables still follow a Gaussian distribution, and sum of Binomial random variables still follow a Binomial distribution. DDP utilizes this nice stability to permit each participant to randomise its local statistic to a lesser degree than would LDP. 
However, in DDP, only the sum of the individually released statistics is $(\epsilon,\delta)$-differentially private but not the individually released statistic, \ie ${\textstyle\sum}{\vec{r_i}}$ is sufficient for this level of differential privacy, but individual noise $\vec{r_i}$ alone is not sufficient, thus $\vec{x_i}+\vec{r_i}$ cannot be released directly. Here $\vec{x_i}$ and ${\vec{r_i}}$ indicate the individual plaintext and noise respectively. 
Therefore, DDP necessitates the help of SMC to maintain utility and ensure aggregator obliviousness, as evidenced in~\cite{rastogi2010differentially,shi2011privacy,acs2011have,lyu2018ppfa,agarwal2018cpsgd,truex2018hybrid,lyu2020lightweight}.

An illustration of one round of FL without privacy and with different DP mechanisms is given in Fig.~\ref{fig:FL_DP}. Another parallel line of work for privacy-preserving distributed learning is to transfer the knowledge of the ensemble of multiple models to a student model~\cite{hamm2016learning,papernot2017semi,papernot2018scalable,lyu2020differentially_kd}. For example, Hamm \etal~\cite{hamm2016learning} first created labeled data from auxiliary unlabeled data, and then used the labeled auxiliary data to find an empirical risk minimizer, finally released a differentially private classifier using output perturbation~\cite{chaudhuri2011differentially}. Similarly, Papernot \etal~\cite{papernot2017semi,papernot2018scalable} proposed \emph{Private Aggregation of Teacher Ensembles} (PATE) to first train an ensemble of teachers on disjoint subsets of private data, then perturb the knowledge of the ensemble of teachers by adding noise to the aggregated teacher votes before transferring the knowledge to a student. Finally, a student model is trained on the aggregate output of the ensemble, such that the student learns to accurately mimic the ensemble. PATE requires a lot of participants to achieve reasonable accuracy, and each participant needs to have enough data to train an accurate model, which might not hold in FL system, where the data distribution of participants might be highly unbalanced, making this approach unsuitable to FL system.

\section{Poisoning Attacks}
\label{sec:Poisoning}
Different from privacy attacks that are targeting at data privacy, poisoning attacks aim to compromise the system robustness. Depending on the attacker's objective, poisoning attacks can be broadly classified into two categories: 1) untargeted poisoning attacks~\cite{Benjamin2009,biggio2012poisoning,jagielski2018manipulating,xie2020fall,fang2020local}; and 2) targeted poisoning attacks~\cite{nelson2008exploiting,huang2011adversarial,bagdasaryan2018backdoor,bhagoji2018analyzing,chen2017targeted,gu2017badnets,shafahi2018poison,liu2017trojaning}. 

\begin{figure}[!htp]
\centering 
\includegraphics[width=1\columnwidth]{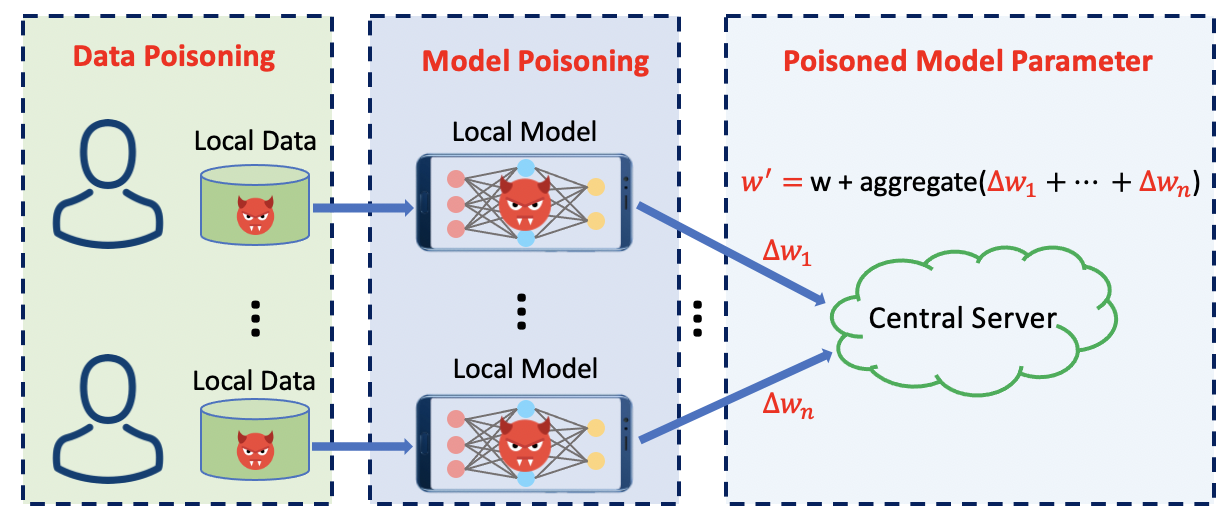}
\caption{Data v.s. model poisoning attacks in FL.}
\label{fig:poisoning_flow1}
\end{figure}

Note that the untargeted and targeted poisoning attacks during the training phase can be mounted on both the data and the model. Fig.~\ref{fig:poisoning_flow1} shows that the poisoned updates can be sourced from two poisoning attacks: (1) data poisoning attack during local data collection; and (2) model poisoning attack during local model training process. At a high level, both poisoning attacks attempt to modify the behavior of the target model in some undesirable way. However, due to the model sharing nature of FL with homogeneous architectures, data poisoning attacks are generally less effective than model poisoning attacks \cite{bagdasaryan2018backdoor,bhagoji2018analyzing,fung2020limitations,sun2019can}. In fact, model poisoning subsumes data poisoning in FL settings, as data poisoning attacks eventually change a subset of updates sent to the model at any given iteration. This is functionally identical to a centralized poisoning attack in which a subset of the training data is poisoned. 

%%%%%%%%%%%%%%%%%%%%%%%%%%%%%%%%%%%%
%%%%%%%%%%%%%%%%%%%%%%%%%%%%%%%%%%%%

\subsection{Untargeted Attacks} 
Untargeted poisoning attacks aim to arbitrarily compromise the integrity of the target model. Byzantine attack is one type of \emph{untargeted} poisoning attacks that uploads arbitrarily malicious gradients to the server so as to cause the failure of the global model~\cite{lamport1982byzantine,blanchard2017machine,damaskinos2019aggregathor,bernstein2019signsgd,xie2020fall}. 
A formal definition of Byzantine attack is given in Definition~\ref{def:byz}.

\begin{definition} 
\label{def:byz}
[Byzantine attack]~\cite{blanchard2017machine,yin2018byzantine}. An honest participant uploads $\Delta\boldsymbol{w}_i\coloneqq\nabla F_i(\boldsymbol{w}_i)$ while a dishonest participant can upload arbitrary values. 
\end{definition}

\begin{align}
\Delta\boldsymbol{w}_i=
\begin{cases}
*, & \mbox{if $i$-th participant is Byzantine}, \\
\nabla F_i(\boldsymbol{w}_i), & \mbox{otherwise,}
\end{cases}
\end{align}
where ``$*$" represents arbitrary values, $F_i$ represents participant~$i$'s local model objective function.

Blanchard \etal~\cite{blanchard2017machine} showed that the aggregation of FL can be completely controlled by a single Byzantine participant if there is no defense in the FL. In particular, suppose there are $n-1$ benign participants and a Byzantine participant, the server aggregates the gradients by $\Delta\boldsymbol{w}'=\frac{1}{n}\sum_{i=1}^n \Delta\boldsymbol{w}_i$, where $\Delta\boldsymbol{w}'$ is the aggregated gradient.
Assume the $n$-th participant is Byzantine, it can always make the aggregated gradient become any vector $\boldsymbol{u}$ by uploading the following gradient:
\begin{equation}
\Delta\boldsymbol{w}_n=n\boldsymbol{u}-\sum_{i=1}^{n-1}\Delta\boldsymbol{w}_i. 
\end{equation}
Such a simple attack exposes the vulnerability of FL against Byzantine attack.

Chen \etal~\cite{chen2020robust} discussed Byzantine attacks in Adam-based FL and proposed a camouflage attack that can camouflage the model updates and launch effective attacks. Their proposed attack also works on other well-known optimizers such as AdaGrad and RMSProp. 
Baruch \etal~\cite{baruch2019little} showed that the core part of gradient descent algorithms is the direction of the descent. Specifically, for gradient descent algorithms, to guarantee the descent of the loss, the inner product between the ground-truth gradient and the robust aggregated gradient must be non-negative:
\begin{equation}
    \begin{split}
    \langle\Delta\boldsymbol{w},\Delta\boldsymbol{w}' \rangle\ge 0,
    \end{split}
\end{equation}
where $\langle\cdot,\cdot \rangle$ is the inner product operation, $\Delta\boldsymbol{w}$ is the optimal gradient, and $\Delta\boldsymbol{w}'=$ aggregate$(\Delta\boldsymbol{w}_1,\cdots,\Delta\boldsymbol{w}_n)$ is the aggregated gradient with aggregate$(\cdot)$ being arbitrary aggregation function.
To make the aggregation fail, they proposed an ``inner product manipulation attack" that can make the inner product between the ground-truth gradient and the robust aggregated gradient negative. To do this, each Byzantine participant uploaded the negative of the average benign gradients. They showed the proposed attack can successfully bypass Coordinate-wise Median~\cite{yin2018byzantine} and Krum~\cite{blanchard2017machine}. Xie \etal~\cite{xie2020fall} claimed that by consistently applying small changes to many parameters, a Byzantine participant can perturb the model’s convergence. First, they used the local data of Byzantine participants to estimate the mean and standard deviation of the distribution.
Then, they analyzed the range in which changes to the parameters will not be detected by the defense, and upon choosing the maxima of this range the convergence is averted. 

\subsection{Targeted Attacks}
In targeted poisoning attacks, the learnt model outputs the target label specified by the adversary for particular testing examples, \eg predicting spams as non-spams, and predicting attacker-desired labels for testing examples with a particular Trojan trigger (backdoor/trojan attacks). However, the testing error for other testing examples is unaffected. 
Generally, targeted attacks is more difficult to conduct than untargeted attacks as the attacker has a specific goal to achieve.

One common example of targeted poisoning attack is the label-flipping attack~\cite{biggio2012poisoning,fung2020limitations}. The labels of honest training examples of one class are flipped to another class while the features of the data are kept unchanged. For example, the malicious participants in the system can poison their dataset by flipping all 1s into 7s. A successful attack produces a model that is unable to correctly classify 1s and incorrectly predicts them to be 7s. 

Another realistic targeted poisoning attack is backdoor poisoning attack, in which an adversary can modify individual features or small regions of the original training dataset to implant a backdoor trigger into the model. The model will behave normally on clean data, yet will constantly predict a target class whenever the trigger (\eg a stamp on an image) appears. For instance, a backdoor attack can cause the FL model to reach 100\% accuracy on the backdoor task, \eg to control an image classifier to assign an attacker-chosen label to images with certain features in an image-classification task, or a next-word predictor completes certain sentences with an attacker-chosen word in a word-prediction task~\cite{bagdasaryan2018backdoor}.

Backdoor attacks can be further divided into two categories: dirty-label attacks~\cite{gu2017badnets,chen2017targeted,liu2020reflection,nguyen2020input} and clean-label attacks~\cite{munoz2017towards,koh2017understanding,shafahi2018poison,zhao2020clean,liu2020reflection,sun2021data}. 
Clean-label attacks assume that the adversary cannot change the label of any training data as there is a process by which data are certified as belonging to the correct class and the poisoning of data samples has to be imperceptible. In contrast, in dirty-label poisoning, the adversary can introduce a number of data samples that are expected to be misclassified by the model with the desired target label into the training data. Clean-label attacks are arguably stealthier as they do not change the labels.

The targeted poisoning attack in FL can be carried out by any FL participant or via collusion on either the data or the gradients. Bhagoji \etal~\cite{bhagoji2018analyzing} demonstrated a single, non-colluding malicious participant can cause the model to misclassify a set of chosen inputs with high confidence. Bagdasaryan \etal~\cite{bagdasaryan2018backdoor} pointed out that the poisoned updates can be generated by training the local model on backdoored local training data, and even a single-shot attack may be enough to inject a backdoor into the global model. Xie \etal ~\cite{xie2020} demonstrated that a global trigger pattern can be decomposed into separate local patterns and embedded into the training set of colluding adversarial participants respectively. The impact on the FL model depends on the extent to which the backdoor participants engage in the attacks, and the amount of training data being poisoned. A recent work shows that poisoning edge-case (low probability) training samples are more effective \cite{wang2020attack}. 

Lastly, we remark that most of the previous research on poisoning attacks focus on Byzantine or backdoor attackers. A system that allows participants to join and leave is susceptible to Sybil attacks~\cite{douceur2002sybil}, in which an attacker gains influence by joining a system to inject $c$ fake participants into the FL system or compromise $c$ benign participants~\cite{fung2020limitations}. Sybil attacks can be launched in both the untargeted and targeted manner. For example, targeted poisoning can be conducted by sybil clones, who contribute updates towards a specific poisoning objective~\cite{fung2020limitations}. Concretely, \cite{fung2020limitations} considered two types of targeted attacks by sybil clones: label-flipping attacks and backdoor attacks. 

%%%%%%%%%%%%%%%%%%%%%%%%%%%%%%%%%%%%%%%%%%%%%%%%%
\begin{table*}[ht]
\caption{The state-of-the-art defenses against federated learning poisoning. $n$ is number of participants. Note that some defenses have no theoretic breaking point.}
\label{tbl:poisoningdefenses_comparison}
\centering
\scalebox{0.9}{
\begin{tabular}{|c|c|c|c|c|c|c|}
\hline
\textbf{Defense} & \textbf{Technique} & \textbf{IID Data} & \textbf{Non-IID Data} 
&\textbf{Breaking Point} 
&\textbf{Targeted Poisoning} &\textbf{Untargeted Poisoning}
\\
\hline
AUROR~\cite{shen2016uror} & Clustering  & $\checkmark$ & $\times$ 
& NA &$\times$ &$\checkmark$ 
 \\ 
\hline
Krum/Multi-Krum~\cite{blanchard2017machine} & Euclidean distance  & $\checkmark$ & $\times$ & $(n-2)/2n$ &$\times$ & $\checkmark$ 
 \\
\hline
Coordinate-wise Median~\cite{yin2018byzantine} & Coordinate-wise median & $\checkmark$ & $\times$ & 1/2 &$\times$ &$\checkmark$ 
 \\
 \hline
 Bulyan~\cite{mhamdi2018hidden} & Krum + trimmed median & $\checkmark$ & $\times$ & $(n-3)/4n$ &$\times$ &$\checkmark$ 
 \\
 \hline
RFA~\cite{pillutla2019robust} & Geometric median & $\checkmark$ & $\times$ & NA &$\times$ &$\checkmark$
\\ 
\hline
FoolsGold~\cite{fung2020limitations} & Contribution similarity & $\checkmark$ & $\checkmark$ & NA &$\checkmark$ &$\times$  
\\ 
\hline
Sun \etal~\cite{sun2019can} &Norm-bounding and DP  &$\checkmark$  &$\checkmark$  &NA  &$\checkmark$ &$\checkmark$~\cite{shejwalkar2021back}  
\\ 
\hline
Wu \etal~\cite{wu2020mitigating} &Pruning  &$\checkmark$  &$\checkmark$ &NA  &$\checkmark$ &$\times$  
\\ 
\hline
CRFL~\cite{xie2021crfl} &Clipping and smoothing  &$\checkmark$  &$\checkmark$ &NA  &$\checkmark$ &$\times$  
\\ 
\hline
\end{tabular}}
\end{table*}

\section{Defenses against Poisoning Attacks}
\label{sec:Poisoning_Defenses}
Robustness to poisoning attacks is a desirable property in FL. To address poisoning attacks, many robust aggregation schemes are proposed in the literature. Known defenses to poisoning attacks in a centralized setting, such as robust losses~\cite{Han2016} and anomaly detection~\cite{Benjamin2009}, assume control of the participants or explicit observation of the training data. Neither of these assumptions are applicable to FL in which the server only observes model parameters/updates sent as part of the iterative ML algorithm~\cite{fung2020limitations}. We summarize the robustness-focused FL defenses against untargeted and targeted attacks as follows.

\subsection{Defenses against Untargeted Attacks}
For Byzantine-resilient aggregation, an algorithm is Byzantine fault tolerant~\cite{blanchard2017machine} if its convergence is robust even when a large portion of participants are adversarial. Below, we list several representative attempts that try to defend against the untargeted Byzantine attacks.

Shen \etal~\cite{shen2016uror} introduced a statistical mechanism called AUROR to detect the malicious users while generating an accurate model. AUROR is based on the observation that indicative features (most important model features) from the majority of honest users will exhibit a similar distribution while those from malicious users will exhibit an anomalous distribution. It then uses k-means to cluster participants' updates across training rounds and discards the outliers, \ie contributions from small clusters that exceed a threshold distance are removed. The accuracy of a model trained using AUROR drops by only 3\% even when 30\% of all the users are adversarial.

Blanchard \etal proposed Krum~\cite{blanchard2017machine}, in which, the top $f$ contributions to the model that are furthest from the mean participant contribution are removed from the aggregation. Krum uses the Euclidean distance to determine which gradient contributions should be removed, and can theoretically withstand poisoning attacks of up to 33\% adversaries in the participant pool, \ie given $n$ agents of which $f$ are Byzantine, Krum requires that $n \geq 2f+3$. At any time step $t$, updates $\{\delta_1^t,\cdots,\delta_n^t\}$ are received at the server. For each $\delta_i^t$, the $n-f-2$ closest (in terms of $L_2$ norm) other updates are chosen to form a set $C_i$, and their distances are added up to give a score $S(\delta_i^t)={\textstyle\sum}_{\delta \in C_i} \|\delta_i^t-\delta\|$. Krum then chooses $\delta_{krum} = \delta_i^t$ with the lowest score to update the global parameter $w_{i}^{t+1}=w_{i}^{t}+\delta_{krum}$. Krum is resistant to attacks by omniscient adversaries – aware of a good estimate of the gradient – who send the opposite vector multiplied by a large factor. It is also resistant to attacks by adversaries who send random vectors drawn from a Gaussian distribution (the larger the variance of the distribution, the stronger the attack). 
Multi-Krum is a variant of Krum, which intuitively interpolates between Krum and averaging, thereby combining the resilience properties of Krum with the convergence speed of averaging. Essentially, Krum filters outliers based on the entire update vector, but does not filter coordinate-wise outliers. 

To address this issue, 
Yin \etal~\cite{yin2018byzantine} proposed two robust distributed gradient descent algorithms, one based on coordinate-wise median, and the other based on the coordinate-wise trimmed mean. Unfortunately, median-based rules can incur a prohibitive computational overhead in large-scale settings~\cite{chen2018draco}. Mhamdi \etal~\cite{mhamdi2018hidden} proposed a meta-aggregation rule called Bulyan, a two-step meta-aggregation algorithm based on the Krum and trimmed median, which filters malicious updates followed by computing the trimmed median of the remaining updates. Median and geometric-median based robust aggregation rules are also extensively explored in~\cite{xie2018generalized,alistarh2018byzantine}. 
Pillutla \etal~\cite{pillutla2019robust} proposed a robust aggregation approach called RFA by replacing the weighted arithmetic mean with an approximate geometric median, so as to reduce the impact of the contaminated updates. Unfortunately, RFA can only handle a few types of poisoning attackers, but not applicable to Byzantine attacks. 

In spite of their robustness guarantees, recent inspections revealed that previous Byzantine-robust FL mechanisms are also quite brittle and can be easily circumvented. Bhagoji \etal~\cite{bhagoji2018analyzing} showed that targeted model poisoning of deep neural networks is effective even against the Byzantine-robust aggregation rules such as Krum and coordinate-wise median. \cite{baruch2019little,xie2020fall} showed that while the Byzantine-robust aggregation rules may ensure that the influence of the Byzantine workers in any single round is limited, the attackers can couple their attacks across the rounds, moving weights significantly away from the desired direction and thus achieve the goal of lowering the model quality. Xu \etal~\cite{xu2020towards} demonstrated that Multi-Krum is not robust against the untargeted poisoning. This is because Multi-Krum is based on the distance between each gradient vector and the mean vector, while the mean vector is not robust against untargeted poisoning. Fang \etal~\cite{fang2020local} showed that aggregation rules (\eg Krum~\cite{blanchard2017machine}, Bulyan~\cite{mhamdi2018hidden}, trimmed mean~\cite{yin2018byzantine}, coordinate-wise median~\cite{yin2018byzantine}, and other median-based aggregators~\cite{chen2017distributed}) that were claimed to be robust against Byzantine failures are not effective in practice against optimized local model poisoning attacks that carefully craft local models on the compromised participants such that the aggregated global model deviates the most towards the inverse of the direction along which the global model would change when there are no attacks. All these highlight the need for more effective defenses against Byzantine attackers in FL.

Other works investigate Byzantine robustness from different lens. Chen \etal~\cite{chen2018draco} presented DRACO, a framework for robust distributed training via algorithmic redundancy. DRACO is robust to arbitrarily malicious computing nodes, while being orders of magnitude faster than state-of-the-art robust distributed systems. However, DRACO assumes that each participant can access other participants' data, limiting its practicability in FL.
Su \etal~\cite{su2018securing} proposed to robustly aggregate the gradients computed by the Byzantine participants based on the filtering procedure proposed by Steinhardt \etal~\cite{steinhardt2017resilience}. Bernstein \etal~\cite{bernstein2019signsgd} proposed {\scriptsize SIGN}SGD, which is combined with majority vote to enable participants to upload element-wise signs of their gradients to defend against three types of half “blind” Byzantine adversaries: (i) adversaries that arbitrarily rescale their stochastic gradient estimate; 
(ii) adversaries that randomise the sign of each coordinate of the stochastic gradient;
(iii) adversaries that invert their stochastic gradient estimate.

\subsection{Defenses against Targeted Attacks}
Existing defenses against the targeted backdoor attacks can be categorized into two types: detection methods and erasing methods~\cite{li2020backdoor}. Detection methods exploit activation statistics or model properties to determine whether a model is backdoored~\cite{wang2019neural,chen2019deepinspect}, or whether a training/test example is a backdoor example~\cite{tran2018spectral}. 

There are a number of detection algorithms that are designed to detect which inputs contain backdoor, and which parts of the model (its activation functions specifically) are responsible for triggering the adversarial behavior of the model, in order to remove the backdoor~\cite{chen2018detecting,chen2017targeted,liu2018fine,liu2017neural,tran2018spectral}. These algorithms rely on the statistical difference between the latent representations of backdoor-enabled and clean (benign) inputs in the poisoned model. These backdoor detection algorithms can however be bypassed by maximizing the latent indistinguishability of backdoor-enabled adversarial inputs and clean inputs~\cite{tan2019bypassing}.

While detection can help identify potential risks, the backdoored model still needs to be purified since the potential impact of backdoor triggers remains uncleared in the backdoored models. The erasing methods take a step further and aim to purify the adverse impacts on models caused by the backdoor triggers. The current state-of-the-art erasing methods are Mode Connectivity Repair (MCR)~\cite{zhao2020bridging} and Neural Attention Distillation (NAD) \cite{li2021neural}. MCR mitigates the backdoors by selecting a robust model in the path of loss landscape, while NAD leverages knowledge distillation to erase triggers. Other previous methods, including finetuning, denoising, and fine-pruning~\cite{liu2018fine}, have been shown to be insufficient against the latest attacks~\cite{yao2019latent,liu2020reflection}. Another more recent work called Anti-Backdoor Learning (ABL) \cite{li2021anti} aims to train clean models given backdoor-poisoned data. They frame the overall learning process as a dual-task of learning the clean and the backdoor portions of data. From this view, they identify two inherent characteristics of backdoor attacks as their weaknesses: 1) the models learn backdoored data much faster than learning with clean data, and the stronger the attack the faster the model converges on backdoored data; 2) the backdoor task is tied to a specific class (the backdoor target class). Based on these two weaknesses, ABL introduces a two-stage gradient ascent mechanism for standard training to 1) help isolate backdoor examples at an early training stage, and 2) break the correlation between backdoor examples and the target class at a later training stage. Extensive experiments on multiple benchmark datasets against 10 state-of-the-art attacks empirically show that ABL can automatically prevent backdoor attacks during training, without degrading the main performance.

Despite the promising backdoor defense results in the centralized setting, it is still unclear whether these defenses can be smoothly adapted to FL setting, especially in the non-iid setting. For backdoor defense in FL, Sun \etal~\cite{sun2019can} showed that clipping the norm of model updates and adding Gaussian noise can mitigate backdoor attacks that are based on the model replacement paradigm. Andreina \etal~\cite{andreina2021baffle} incorporated an additional validation phase in each round of FL to detect backdoor.
However, None of these provides certified robustness guarantees. Certified robustness for FL against backdoor attacks remain largely unexplored. Xie \etal~\cite{xie2021crfl} provided the first general framework called \emph{Certifiably Robust Federated Learning} (CRFL), to train certifiably robust FL models against backdoors.

To defend against the targeted poisoning attack by sybil clones, Fung \etal~\cite{fung2020limitations} exploited the characteristic behavior that sybils are more similar to each other than the similarity observed amongst the honest clients, and proposed FoolsGold: a new defense scheme against FL Sybil attacks by adapting the learning rate of participants based on contribution similarity. Note that FoolsGold does not bound the expected number of attackers by assuming that attackers can spawn a large number of Sybils, rendering assumptions about proportions of honest participants unrealistic~\cite{blanchard2017machine}. Additionally, FoolsGold requires no auxiliary information beyond the learning process, and makes fewer assumptions about participants and their data. The robustness of FoolsGold holds for different distributions of participant data, varying poisoning targets, and various Sybil strategies, and can be applied successfully on both FedSGD and FedAvg.

We list the most representative defenses against poisoning attacks against FL in Table~\ref{tbl:poisoningdefenses_comparison}. Some robust aggregation algorithms have breaking points, \ie the fraction of malicious participants, robustness guarantees cannot be provided if the fraction of malicious participants is larger than the breaking point. 

\textbf{Remark}. Note that both the untargeted and targeted poisoning attacks are less effective in settings with
infrequent participation like H2C \cite{bagdasaryan2018backdoor}. Moreover, under practical production FL environments, Shejwalkar \etal~\cite{shejwalkar2021back} have showed that FL, even without any defenses, is highly robust in practice. For production cross-device FL (H2C), which contains thousands to billions of clients, poisoning attacks have no impact on existing robust FL algorithms even with impractically high percentages of compromised clients. For production cross-silo FL (H2B), which contains up to hundred clients, data poisoning attacks are completely ineffective; model poisoning attacks are unlikely to play a major risk when the clients involved are bound by contract and their software stacks professionally maintained (e.g., in banks, hospitals). Some exceptional cross-silo scenarios are most likely with a strong incentive (e.g., financial) causing multiple parties to be willing to risk breach of contract by colluding or for one party to hack thereby risking criminal liability. Therefore, we conclude that these poisoning attacks are more likely to happen in some exceptional H2B scenarios.

\section{Discussions and Promising Directions}
\label{sec:future}
There are still potential vulnerabilities which need to be addressed in order to improve the privacy and robustness of FL systems. Moreover, there are multiple design goals that are equally important with privacy and robustness, thus need to be considered simultaneously in FL. In this section, we outline research directions which we believe are promising.

\textbf{Curse of Dimensionality:}
Large models, with high dimensional parameter vectors, are particularly susceptible to privacy and security attacks~\cite{chang2019cronus}. Most FL algorithms require overwriting the local model parameters with the global model. This makes them susceptible to poisoning attacks, as the adversary can make small but damaging changes in the high-dimensional models without being detected. Almost all of the well-designed Byzantine-robust aggregators~\cite{blanchard2017machine,yin2018byzantine,pillutla2019robust} still suffer from the curse of dimensionality. Specifically, the estimation error scales up with the size of the model in a square-root manner. Thus, sharing model parameters may not be a strong design choice in FL, it opens all the internal state of the model to inference attacks, and maximizes the model's malleability by poisoning attacks. The large number of hyperparameters might also adversely affect communication and accuracy, though the follow up work~\cite{thakkar2019differentially} tried adaptive gradient clipping strategies to help alleviate this issue. To address these fundamental shortcomings of FL, it is worthwhile to explore whether sharing gradients is essential. Instead, sharing less sensitive information (\eg {\scriptsize SIGN}SGD~\cite{bernstein2018signsgd}) or only sharing model predictions~\cite{chang2019cronus,li2019fedmd,sun2020federated} in a black-box manner may result in more robust privacy protection in FL. 

\textbf{Rethinking Current Privacy Attacks:}
There are several inherent weaknesses in current attacks that may limit their applicability in FL~\cite{lyu2021novel}. For example, GAN attack assumes that the entire training corpus for a given class comes from a single participant, and only in the special case where all class members are similar, GAN-constructed representatives are similar to the training data~\cite{hitaj2017deep}. These assumptions may be less practical in FL. For DLG~\cite{zhu2019deep} and iDLG~\cite{zhao2020idlg}, both works: (1) adopt a second-order optimization method called L-BFGS, which is more computationally expensive compared with first-order optimization methods; (2) are only applicable to gradients computed on mini-batches of data, \ie at most B=8 in DLG, and B=1 in iDLG, which is not the real case for FL, in which gradient is normally shared after at least 1 epoch of local training; (3) used untrained model, neglecting gradients over multiple communication rounds. Attacking FL system in a more efficient manner and under more practical settings remains largely unexplored.

\textbf{Rethinking Current Defenses:}
FL with secure aggregation for the purpose of privacy is more susceptible to poisoning attacks as the individual updates cannot be inspected. Similarly, it is still unclear if adversarial training, one state-of-the-art defense approach against adversarial attacks in conventional ML \cite{madry2018towards,wang2019convergence,wang2019improving}, can be adapted to FL, as adversarial training was developed primarily for IID data and remains unclear for its performance in non-IID settings. Moreover, adversarial training is computationally expensive and may hurt the performance \cite{tsipras2018robustness}, which may not be feasible for H2C scenario. In terms of differential privacy (DP) based methods~\cite{mcmahan2018learning,lyu2019fog,zhao2020local,lyu2020towards,lyu2020how,lyu2020foreseen},
record-level DP bounds the success of membership inference, but does not prevent property inference applied to a group of training records~\cite{melis2019exploiting}. Participant-level DP, on the other hand, is geared to work with thousands of users for training to converge and achieving an acceptable trade-off between privacy and accuracy~\cite{mcmahan2018learning}. The FL model fails to converge with a small number of participants, making it unsuitable for H2B scenarios. Furthermore, DP may hurt the accuracy of the learned model~\cite{pan2020privacy}, which may not be appealing to the industry. Further work is needed to investigate if participant-level DP can protect FL systems with fewer participants.

\textbf{Optimizing Defense Mechanism Deployment:}
When deploying defense mechanisms to check if any adversary is attacking the FL system, the FL server will need additional computational cost. In addition, different types of defense mechanisms may have different effectiveness against different attacks, and incur different costs. It is important to study how to optimize the timing of deploying the defense mechanisms or the announcement of deterrence measures. Game theoretic research holds promise in addressing this challenge.

\textbf{Test-phase Privacy in FL:}
This survey mainly focuses on the training phase attacks and defenses in FL, considering the more attack possibilities opened by the distributed training property of FL systems. In fact, FL is also vulnerable to both privacy and robustness attacks during test/inference phase by the users of the final FL model when it is deployed as a service.

In terms of the privacy vulnerability, the trained global model may reveal sensitive information from model predictions when deployed as a service, causing privacy leakage. In such a setting, an adversary does not have direct access to the model parameters, but may be able to view input-output pairs.
Previous researches have shown a series of privacy leakage given only black-box access to the trained models, such as (1) model stealing attacks in which model parameters can be reconstructed by an adversary who only has access to an inference/prediction API based on those parameters~\cite{tramer2016stealing,he2021model,xu2021beyond,chen2021killing}; (2) membership inference attacks which aim to determine if a particular record was used to train the model~\cite{shokri2017membership}. FL models face the similar dilemma during model deployment for testing purpose. The development of effective defenses against privacy leakage during model deployment calls for further investigations.

\textbf{Test-phase Robustness in FL:}
In terms of the robustness vulnerability, recent studies~\cite{zizzo2020fat,hong2021federated,shah2021adversarial} have shown that FL is also vulnerable to well-crafted adversarial examples. During inference time, the attackers can add a very small perturbation to the test data, making the test data almost indistinguishable from natural data and yet classified incorrectly by the global model. For federated robustness against adversarial examples,~\cite{zizzo2020fat,hong2021federated} proposed to apply adversarial training (AT) to FL, i.e, federated adversarial training (FAT), in order to achieve adversarial robustness in FL. 
\cite{zizzo2020fat} noticed that conducting AT on all participants leads to divergence of the model. To solve this problem, they conducted AT on only a proportion of participants for better convergence.
Another recent work~\cite{hong2021federated} considered hardware heterogeneity in FL, \ie only limited users can afford AT. Hence, they conduct AT on only a proportion of participants that have powerful computation resources while conducting standard training on the rest of the participants. 
\cite{shah2021adversarial} investigated the impact of communication rounds in FAT and proposed a dynamic adversarial training.
The training of all the above FAT works are unstable, which potentially hurts the convergence and performance. Moreover, adversarial training typically requires significant computation and longer time to converge, and it is unclear how it performs in non-IID settings. How to speed up adversarial training in FL and the investigation of its applicability in non-IID settings may be required in the future.
Overall, there exist difficulties in applying adversarial training to the federated setting. This motivates future works to explore more effective approaches to maintain both natural accuracy and robust accuracy in FL. 

In addition to the adversarial examples, recent works~\cite{he2021model,xu2021beyond} have validated that the API services (the victim/target model) can be easily stolen and are vulnerable to adversarial example transferability attack. It would be interesting to explore whether the collaboratively built global model in FL is also facing the similar problem, and how to effectively claim the ownership of the trained model~\cite{he2021protecting}. 

Overall, it would be of much importance towards realizing trustworthy FL by defending against both training-phase and test-phase attacks.

\textbf{Relationship with GDPR:}
GDPR \footnote{\url{https://gdpr-info.eu}} defines 6-core principles as rational guidelines for service providers to manage personal data, including: (1) Lawfulness, fairness and transparency; (2) Purpose limitation; (3) Data minimisation; (4) Accuracy; (5) Storage limitation; (6) Integrity and confidentiality (security). GDPR also requires Data Controllers to provide the following rights for Data Subjects if capable (The GDPR Articles 12–23): (1) Right to be informed, (2) Right of access, (3) Right to rectification, (4) Right to erasure (Right to be forgotten), (5) Right to restrict processing, (6) Right to data portability, (7) Right to object, and (8) Rights in relation to automated decision making and profiling.
Although FL has emerged as a prospective solution that facilitates distributed collaborative learning without disclosing original training data, unfortunately, FL is not naturally compliant with the GDPR~\cite{truong2021privacy}, as pointed out by a recent survey~\cite{truong2021privacy} that has dedicated to surveying the relationship between FL and GDPR requirements. For example, secure aggregation mechanism in FL amplifies the lack of transparency and fairness in FL systems, thus fails to fully comply with the GDPR requirements of fairness and transparency; malicious participants in FL may conduct either data or model poisoning attack for an unauthorised purpose, local ML model parameters obtained from participants is no longer minimal for the original purpose. These possible attacks, which lead to non-compliance with the GDPR, should be addressed. Henceforth, it is worthwhile to explore approaches to empower FL-based systems to follow the GDPR regulatory guidelines, thus fully comply with the GDPR.

\textbf{Threats and Protections of VFL and FTL:}
This survey mainly focuses on the threats to HFL, there are some recent exploratory efforts on threats and protections of VFL and FTL.

For VFL, Secureboost~\cite{cheng2021secureboost} considered user privacy and data confidentiality in VFL, and presented an approach to train a high-quality tree boosting model collaboratively. 
A recent work called FederBoost~\cite{tian2020federboost} pointed out that Secureboost is expensive since it requires cryptoraphic computation and communication for each possible split, thus they proposed a vertical FederBoost which does {\em not} require any cryptographic operation. 
Another recent work~\cite{jin2021catastrophic} uncovered the risk of  \emph{catastrophic data leakage in vertical federated learning} (CAFE) through a novel algorithm that can perform large-batch data leakage with high data recovery quality and theoretical guarantees. They empirically demonstrated that CAFE can recover large-scale private data from the shared aggregated gradients in VFL settings, overcoming the batch limitation problem in current data leakage attacks. 

For FTL, \cite{gao2019privacy} proposed an end-to-end privacy-preserving multi-party learning approach with two variants based on homomorphic encryption and secret sharing techniques, respectively, to build a heterogeneous federated transfer learning (HFTL) framework. \cite{liu2020secure} adopted two secure approaches, namely, homomorphic encryption (HE) and secret sharing for preserving privacy. The HE approach is simple but computationally expensive. By contrast, the secret sharing approach offers the following advantages: (i) there is no accuracy loss, (ii) computation is much faster than HE approach. The major drawback of the secret sharing approach is that one has to offline generate and store many triplets before online computation.

Overall, there is still a large space for VFL and FTL. It is worth further investigation as for whether existing threats in HFL are all valid in VFL and FTL, or if there are new threats and countermeasures in VFL and FTL.

\textbf{Vulnerabilities to Free-riding Participants:}
In FL systems, there may exist free-riders, who aim to benefit from the global model, but do not want to contribute any useful information, thus compromising collaborative fairness~\cite{lyu2020towards,lyu2020how,xinyi2021gradient}. The main incentives for free-riders include: (1) the participant dose not have any data to train the local model; (2) the participant is too concerned about its privacy thus chooses to release fake updates; (3) the participant does not want to consume or does not have any local computation power to train the local model. In the current FL paradigm~\cite{mcmahan2017communication}, all participants receive the same federated model at the end of the collaborative training, regardless of their individual contributions. This makes the paradigm vulnerable to free-riding participants~\cite{lyu2020towards,lyu2020how,lyu2020Collaborative,xu2020towards,FLPI2020}. How to prevent free-riding remains an open challenge.

\textbf{More Possibilities in FL with Heterogeneous Architectures:} Most privacy and robustness researches are focused on FL with homogeneous architectures. It remains unclear whether existing attacks, privacy-preserving techniques and defense mechanisms can be adapted to FL with heterogeneous architectures. It is valuable future work to explore similar types of attacks and defenses in heterogeneous FL.

\textbf{Decentralized FL:}
Decentralized FL is an emerging research area, where there is no single central server in the system~\cite{mcmahan2018learning,FL2019,lyu2020towards,lyu2020how}. Decentralized FL is potentially more useful for H2B scenarios where the business participants do not trust any third party. In this paradigm, each participant could be elected as a server in a round robin manner. The recent emerging swarm learning~\cite{warnat2021swarm} can be deemed as a decentralized FL framework, which unites edge computing, blockchain-based peer-to-peer networking and coordination while maintaining confidentiality without the need for a central coordinator. It is interesting to investigate whether existing threats to server-based FL still apply in decentralized FL. 

\textbf{Efficient FL with Single Round Communication:}
In addition to privacy and robustness, communication cost is another major concern that may hinder the practical deployment of FL. One-shot FL has recently emerged as a promising approach for communication efficiency. It allows the central server to learn a model in a single communication round. Despite the low communication cost, existing one-shot FL methods are mostly impractical or face inherent limitations, \eg a public dataset is required, participants' models are homogeneous, additional data/model information needs to be uploaded, unsatisfactory performance~\cite{guha2019one,li2020practical,dennis2021heterogeneity}. A recent work proposed a more practical data-free approach named FedSyn for one-shot FL framework with heterogeneity~\cite{zhang2021practical}. FedSyn is the first method that can be practically applied to various real-world applications due to the following benefits:
(1) FedSyn requires no additional information (except the model parameters) to be transferred between participants and the server;
(2) FedSyn does not require any auxiliary dataset for training;
(3) FedSyn is the first to consider both model and statistical heterogeneities in FL, \ie the participants' data are non-iid and different participants may have different model architectures. Other alternative one-shot FL approaches with practical assumptions are worthwhile to explore, considering the alluring communication efficiency and less  privacy and robustness attack surfaces exposed in one-shot FL.

\textbf{Achieving Multiple Objectives Simultaneously:}
There are no existing works that can satisfy multiple goals simultaneously: (1) fast algorithmic convergence; (2) good generalisation performance; (3) communication efficiency; (4) fault tolerance; (5) privacy preservation; and (6) robustness to targeted, untargeted poisoning attacks, and free-riders. Previous works have attempted to solve multiple objectives at the same time. For example, ~\cite{lyu2020how,lyu2020towards} addressed collaborative fairness and privacy simultaneously; ~\cite{xu2020towards,xu2021reputation} proposed a \emph{Robust and Fair Federated Learning} (RFFL) framework to address both collaborative fairness and Byzantine robustness. However, it is important to highlight that there is an inherent conflict between privacy and robustness: defending against robustness attacks usually require complete control of the training process or access to the training data~\cite{barreno2010security,shen2016uror,blanchard2017machine,yin2018byzantine,fung2020limitations,steinhardt2017certified}, which goes against the privacy requirements of FL. Although using encryption or DP-based techniques can provide provably privacy preservation, they are not robust to poisoning attacks and may produce models with undesirably poor privacy-utility trade-offs.
Agarwal \etal~\cite{agarwal2018cpsgd} combined differential privacy with model compression techniques to reduce communication cost and obtain privacy benefits simultaneously. It remains largely unexplored and there exist large gaps as for how to simultaneously achieve all the above six objectives.

\section{Conclusions}
\label{sec:Conclusions}
Although federated learning is still in its infancy, it will continue to thrive and will be an active and important research area in the foreseeable future.
As FL evolves, so will the privacy and robustness threats to FL. It is of vital importance to provide a broad overview of current attacks and defenses on FL so that future FL system designers are well aware of the potential vulnerabilities in the current designs, and help them clear roadblocks towards the real-world deployment of FL. This survey serves as a concise and accessible overview of this topic, and it would greatly help our understanding of the privacy and robustness attack and defense landscape in FL. Global collaboration on FL is emerging through a number of workshops in leading AI conferences\footnote{\url{http://www.federated-learning.org/}}. The ultimate goal of developing a general purpose FL defense mechanism that can be robust against various attacks without degrading model performance will require interdisciplinary effort from the wider research community.
 
% \ifCLASSOPTIONcompsoc
%   % The Computer Society usually uses the plural form
%   \section*{Acknowledgments}
% \else
%   % regular IEEE prefers the singular form
%   \section*{Acknowledgment}
% \fi

% This research is supported, in part, by the Joint NTU-WeBank Research Centre on Fintech (Award No: NWJ-2020-008), Nanyang Technological University, Singapore; Joint SDU-NTU Centre for Artificial Intelligence Research (C-FAIR) (NSC-2019-011); the National Research Foundation, Singapore under its AI Singapore Programme (AISG Award No: AISG2-RP-2020-019)); the RIE 2020 Advanced Manufacturing and Engineering (AME) Programmatic Fund (No. A20G8b0102), Singapore; and Nanyang Technological University, Nanyang Assistant Professorship (NAP). Qiang Yang is supported in part by the Hong Kong RGC theme-based research scheme (T41-603/20-R). Any opinions, findings and conclusions or recommendations expressed in this material are those of the author(s) and do not reflect the views of National Research Foundation, Singapore.

\bibliographystyle{IEEEtran}
\bibliography{survey_biblio}

\end{document}